\documentclass[10pt,journal,letterpaper,compsoc]{IEEEtran}
\usepackage{epsfig}
\usepackage{amsmath}
\usepackage{latexsym}
\usepackage{amsfonts}
\usepackage{color}
\usepackage{subfigure}
\usepackage{url}
\usepackage{epsfig}
\usepackage{amssymb}
\usepackage[table ]{xcolor}

\usepackage{algorithmic}
\usepackage{algorithm}

\definecolor{gray}{gray}{0.3}

\allowdisplaybreaks[4]

\newtheorem{lemma}{Lemma}

\newtheorem{theorem}{Theorem}

\begin{document}

\title{Throughput Maximization of UAV Networks}
\author{Wenzheng Xu, \IEEEmembership{Member, IEEE}, Yueying Sun, Rui Zou, Weifa Liang, \IEEEmembership{Senior Member, IEEE},  Qiufen Xia, Feng Shan,  Tian Wang,
Xiaohua Jia, \IEEEmembership{Fellow, IEEE},  and Zheng Li
\IEEEcompsocitemizethanks{
 \IEEEcompsocthanksitem
 Wenzheng Xu, Yueying Sun, Rui Zou,
and Zheng Li (corresponding author)  are with College of Computer Science, Sichuan University, Chengdu, 610065, P. R. China.
E-mails:  wenzheng.xu3@gmail.com,  sunyueying1207@icloud.com, 525093672@qq.com, lizheng@scu.edu.cn

\IEEEcompsocthanksitem Weifa Liang
and
Xiaohua Jia are with Department of Computer Science,  City University of Hong Kong, 83 Tat Chee Ave., Kowloon, Hong Kong,
E-mails: wliang@cs.anu.edu.au, csjia@cityu.edu.hk

\IEEEcompsocthanksitem Qiufen Xia is with School of Software, Dalian University of Technology, Dalian, 116024, P. R. China,  E-mail: qiufenxia@dlut.edu.cn

\IEEEcompsocthanksitem Feng Shan is with School of Computer Science and Engineering,
Southeast University, Nanjing, 211189, P. R. China,  E-mail: shanfeng@seu.edu.cn

\IEEEcompsocthanksitem  Tian Wang is with College of Computer Science and Technology, Huaqiao University,  P.R. China. E-mail: cs\_tianwang@163.com
 }
}

\markboth{IEEE/ACM Transactions on Networking,~Vol.~X, No.~X, XX,~2021}
{Xu \MakeLowercase{\textit{et al.}}:
Coverage Maximization of UAV Networks}

\IEEEcompsoctitleabstractindextext{
\begin{abstract}
In this paper we study the  deployment of multiple unmanned aerial vehicles (UAVs) to form a temporal UAV network for the provisioning of  emergent communications to affected people in a disaster zone,
where each UAV is equipped with a lightweight base station device and thus
can act as an aerial base station for users.
Unlike most existing studies that assumed that  a  UAV can serve all users in its communication range,
we observe that both computation and communication capabilities
of a single lightweight UAV are very limited, due to various constraints on its  size, weight, and power supply. Thus, a single UAV can only provide communication services to a limited number of users.
We study  a novel problem of deploying $K$ UAVs in the top of a disaster area
 such that
  the sum of the data rates of users served by the UAVs is maximized, subject to that (i) the number of users served by each UAV is no greater than its service capacity; and (ii) the communication network induced by the $K$ UAVs is connected. We then propose a $\frac{1-1/e}{\lfloor \sqrt{K} \rfloor}$-approximation algorithm for the problem, improving the current best result of the problem by five times (the best approximation ratio so far is $\frac{1-1/e}{5( \sqrt{K} +1)}$), where $e$ is the base of the natural logarithm. We finally evaluate the algorithm performance via simulation experiments. Experimental results show that the proposed algorithm is very promising. Especially,
the solution delivered by the proposed algorithm is up to 12\% better than those by existing algorithms.
\end{abstract}

}

\date{}
\maketitle

\IEEEdisplaynotcompsoctitleabstractindextext
\IEEEpeerreviewmaketitle

\allowdisplaybreaks

\section{Introduction} \label{secIntro}

It is estimated that the annual loss incurred by natural disasters (e.g.,
earthquakes, tsunamis, flooding, etc.) is about US\$115 billion in the past 30 years.
And even worse, on average 48 thousand people died in  disasters per year~\cite{NatCatSERVICE16}.
When a disaster event occurs,  existing communication and transportation
  infrastructures may have been totally destroyed already.
It is well recognized that
the first 72 hours after the disaster are the golden time window for people life rescues, and search and rescue operations must be conducted quickly and efficiently~\cite{ENNA17, LXL19}.
To rescue the people trapped in the disaster zone,
it is urgent to have temporarily emergent communications
to help them get out from there as soon as possible.


\begin{figure}[htp]
\begin{center}
\includegraphics[scale=0.19]{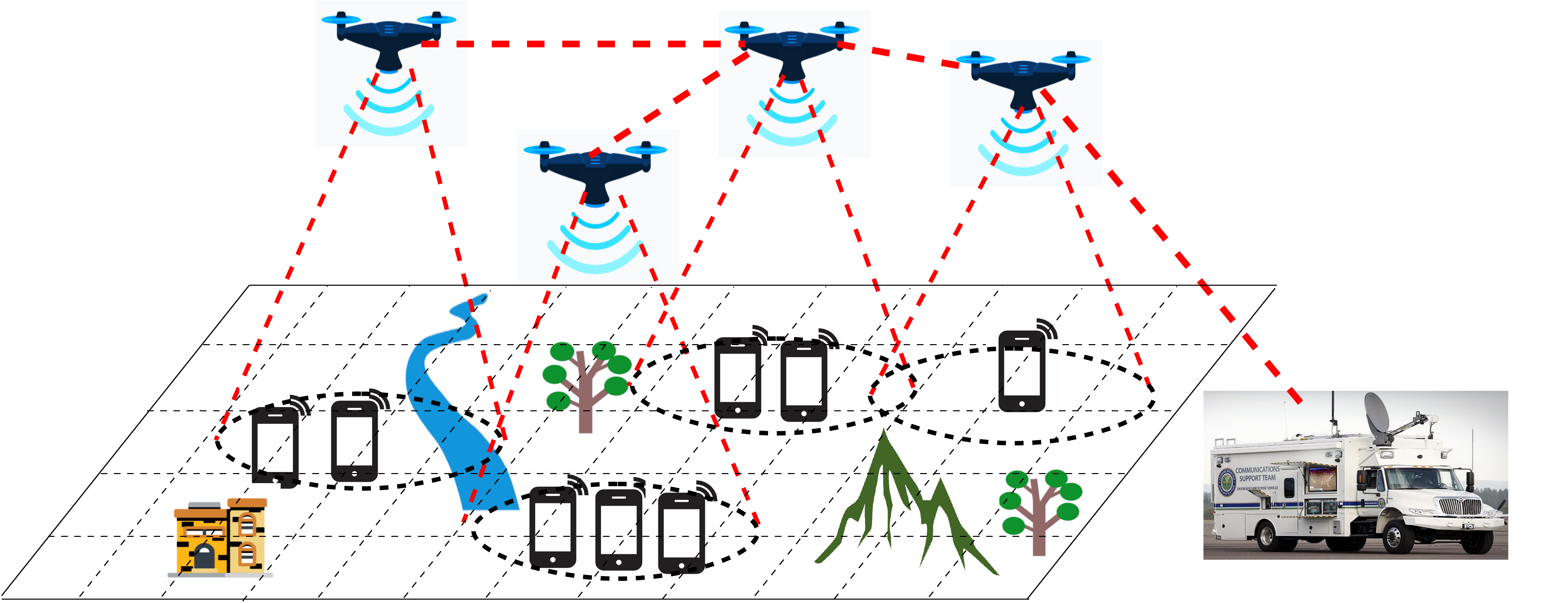}
  \vspace{-4mm}
\caption{A UAV network that provides services for ground users in a disaster area, where the network is connected to the Internet via an emergency communication vehicle.}
 \label{figUAVnetwork}
 \end{center}
 \vspace{-8mm}
\end{figure}

On the other hand, wireless communications by utilizing UAVs recently have attracted a lot of attentions~\cite{ENNA17, LG14, MF15, MTKG16,THMI16, ZZL16, ZLS19}.
Unlike terrestrial communication infrastructures,
low-altitude UAVs are more cost-effective, swift and flexible for on-demand deployments~\cite{ZZ17}, where  UAVs can work as aerial base stations by attaching lightweight base station devices~\cite{CGA+16, MSC+18}. Several mobile operators, including AT\&T and Verizon in the United States, have conducted trials with LTE base stations mounted on UAVs~\cite{MSC+18}.
It is recognized that a UAV network composed of multiple UAVs is perfectly applicable for temporary and unexpected burst communication scenarios, such as natural disaster reliefs, concerts, and traffic congestion~\cite{BY16}, where multiple UAVs can be temporarily  deployed in the top of a disaster area to provide efficient communication services to ground users.
Another advantage of UAV communications is that,
they usually enjoy higher Line-of-Sight (LoS) link opportunity with ground users,
due to  high heights of UAVs, thereby having higher data rates and larger communication ranges~\cite{AKL14}.
Fig.~\ref{figUAVnetwork} illustrates a UAV network with four UAVs, and the UAVs  provide communication services (e.g.,  LTE or WiFi) to the ground people in a disaster area.
With the aid of the UAV network, the people can send and receive emergent data,  such as voice and  video, to/from a rescue team nearby, thereby reducing their injuries and
saving  lives.

The deployment of UAV networks recently has gained lots of attentions~\cite{LCTXP18, SLC+19, YCX+19, ZWWW18}. For example, Zhao {\it et al.}~\cite{ZWWW18} presented a motion control algorithm for deploying  $K$ UAVs to cover as many users as possible while maintaining the connectivity of the $K$ UAVs. Liu {\it et al.}~\cite{LCTXP18} considered a similar problem and proposed a deep reinforcement learning (DRL) based algorithm. Yang {\it et al.}~\cite{YCX+19} investigated the problem of scheduling multiple UAVs to fairly provide communication services to mobile ground users for a given period, by using the DRL method, too.

In spite of the aforementioned studies on the deployment of UAVs,
most of them did not consider the service capacities of the UAVs and
assumed that each UAV can serve all users in its communication range.
On the other hand, due to the constraint on the payload of a UAV,
e.g., the maximum payload a  DJI Matrice M300 RTK UAV is only 2.7~kg~\cite{M300-15},
the computation capacity of the base station device
mounted on the UAV
is very limited~\cite{CYA+18, CGA+16, MSC+18}.
Then, the service capacity
of each UAV, i.e., the maximum number of users that the UAV can serve, is very limited.
Furthermore, it is shown that users may not be uniformly distributed in a monitoring area.
It is very likely that there are many people at a small portion of places while only a few people at the other places. That is, the human density follows the power law~\cite{SKWB10}. For example, after an earthquake, many people may stay in a public plaza without surrounding buildings for their safeties.
It can be seen that only one UAV may be deployed in the top of a dense location with many people by the existing studies~\cite{LCTXP18,  YCX+19, ZWWW18},
as the people are within the transmission range of the UAV.
However, only a portion of the people can be served by the UAV
due to its limited service capacity. A simple solution to address this issue is to deploy multiple UAVs to cover such a dense location with many people. However, a difficult challenge lies in that there may be multiple such dense
locations in a disaster area which are far away from each other, while the communication network formed by the deployed UAVs at different locations may not be connected.

In this paper, we consider that each UAV can provide communication services to  limited numbers of users simultaneously. We study a novel {\it connected maximum  throughput problem}, which is to deploy $K$ UAVs for serving people in a disaster zone, such that the sum of the data rates  of users served is maximized, subject to that (i) the number of users served by each UAV is no greater than its service capacity; and (ii) the communication subnetwork induced by the $K$ UAVs is connected.
Tackling this defined problem poses the following challenges:

(1) Among potential hovering locations for the UAVs, which $K$ locations should be chosen such that the sum of the data rates of users is maximized, subject to that each UAV has a limited service capacity. (2) How to ensure that the communication network induced by the $K$ UAVs at the $K$ identified locations is connected. (3) Consider that the problem is NP-hard, developing
an approximation algorithm with a provable approximation ratio for the problem is extremely difficult, as not only do we need finding the $K$ locations for the $K$ UAVs to maximize the sum of user data rates, but also we must ensure that the communication network induced by the $K$ UAVs is connected.

The novelty of this paper lies in not only incorporating the service capacities of UAVs into consideration but also proposing a performance-guaranteed approximation algorithm to the connected maximum throughput problem in a UAV network.
 Specifically, the proposed algorithm delivers a $\frac{1-1/e}{\lfloor \sqrt{K} \rfloor}$-approximate solution to the problem, which significantly improves the best $\frac{1-1/e}{ 5(\sqrt{K} +1)}$-approximate solution so far by five times~\cite{KLT15}, where $K$ is the number of UAVs, $e$ is the base of the natural logarithm.


The main contributions of this paper can be summarized as follows.
We first formulate a novel connected maximum throughput problem for a UAV network that consists of $K$ UAVs. We then devise a $\frac{1-1/e}{\lfloor \sqrt{K} \rfloor}$-approximation algorithm for the problem. Particularly, the approximation ratio of the algorithm is only $\frac{1-1/e}{7}$ when $K$ (=50) UAVs are deployed. We finally evaluate the proposed algorithm performance via  simulation experiments. Experimental results show that the algorithm is very promising. Especially, the sum of the data rates of users served by the proposed algorithm is up to 12\% larger than those by existing algorithms.

The rest of this paper is organized as follows.
Section~\ref{secRelatedWork} reviews related work.
Section~\ref{secPrelim} introduces the system model and defines the problem precisely.
Section~\ref{secMaxAssign} proposes an efficient algorithm for a subproblem of the connected maximum throughput problem that will serve as a subroutine of the proposed approximation algorithm. Section~\ref{secApprAlg} devises an approximation algorithm for the problem, and Section~\ref{secAlgAnalysis} analyzes the approximation ratio of the approximation algorithm. Section~\ref{secPerformance} evaluates the performance of the proposed algorithm via simulation experiments, and Section~\ref{secCon} concludes this paper.

\section{Related Work} \label{secRelatedWork}

The utilization of UAVs as aerial base stations  has attracted a lot of attentions recently.
For example, Zhao {\it et al.}~\cite{ZWWW18} proposed a distributed
motion control algorithm to deploy a fixed number $K$ of UAVs to serve
as many as users,  while guaranteeing the connectivity of the UAV network.
Liu {\it et al.}~\cite{LCTXP18}
designed  an improved algorithm for  a similar problem, by adopting a
deep reinforcement learning.
Yang {\it et al.}~\cite{YCX+19} studied a problem of scheduling  multiple UAVs to  provide communications to  users within a  period in a fair way, by using the deep reinforcement method, too.
Shi {\it et al.}~\cite{SLC+19} investigated a problem of finding the
flying tours of  UAVs for a period so that
the average UAV-to-user pathloss  is minimized.
They decoupled the problem into several local optimization subproblems,
and solved the subproblems independently.
Selim {\it et al.}~\cite{SK18} proposed a UAV network infrastructure,
which consists of  three types of UAVs: tethered backhaul
UAVs that provide high capacity backhauling, untethered
communication UAVs that provide communication service to ground users,
and untethered powering UAVs for charging communication UAVs.
They studied a problem of finding the placement locations of UAVs
such that  the energy
consumption of the UAVs is minimized, where each
user has  a minimum data rate.
Also, Alzidaneen {\it et al.}~\cite{AAA20} considered a network that consists of UAVs and tethered balloons,
where UAVs can communicate with ground users while UAVs
access the Internet by directly communicating with the balloons.
It can be seen that the coverage area of the network is limited,
since UAVs must be within the communication range of the tethered
balloons. They studied a problem of finding the hovering locations of
UAVs, the association between UAVs and users, and the association
between UAVs and balloons, such that the sum of the data rates of
all users is maximized.
However, most of these mentioned studies did not take the service capacity of
each UAV into consideration, and assumed that a UAV can
serve all users in its communication range.
On the other hand, it is not uncommon that both the
computation and communication capabilities
of a UAV are limited, due to various constraints on its  size, weight, and power supply~\cite{CYA+18, CGA+16}.
In this paper, we  assumed that each
 UAV can provide communication services to  limited numbers of users simultaneously.

On the other hand, there are several studies on maximizing submodular functions
subject to the connectivity constraint.
For example,
Kuo~{\it et al.}~\cite{KLT15} considered the problem of deploying $K$ wireless routers in a network such that a submodular function of the deployed routers is maximized, subject to the connectivity constraint that the subnetwork induced by the $K$ routers is connected.
They proposed a $\frac{1-1/e}{5(\sqrt{K}+1)}$-approximation algorithm,
which is inferior to the approximation ratio $\frac{1-1/e}{\lfloor \sqrt{K} \rfloor}$ of the algorithm obtained in this paper,
where $e$ is the base of the natural logarithm. Khuller~{\it et al.}~\cite{KPK14, KPS20} studied a  problem
of choosing $K$ nodes in a network so that a {\it special submodular function} of the
chosen nodes is maximized,
subject to the connectivity constraint,
where $f$ is a special submodular if (i) $f$ is submodular; and
(ii) $f(A \cup B) = f(A )+f(B)$
if $N(A) \cap N(B) = \emptyset$ for any $A, B\subset V$,
and $N(X)$ denotes the neighborhood of a set $X$, including $X$ itself.
They designed a $\frac{1-1/e}{12}$-approximation algorithm.
However, the objective function of the  problem considered in this paper
may not be a special submodular function.
In addition,  Huang~{\it et al.}~\cite{HLS15}
investigated the  problem of placing $K$ sensors to monitor targets
so that the number of targets covered by the $K$ sensors is maximized
and the network formed by the $K$ placed sensors is connected,
where a target is covered by a sensor if their Euclidean distance is no more than
a given sensing range $R_s$, and two sensors can communicate with each
other if their Euclidean distance is no greater than a given communication range $R_c$, and $R_s\le R_c$. They proposed a  $\frac{1-1/e}{8(\lceil 2\sqrt{2} \alpha \rceil+1)^2}$-approximation algorithm, where $\alpha=\frac{R_s}{R_c}$,
and the ratio thus is between $\frac{1-1/e}{128}$ and $\frac{1-1/e}{32}$, as $0 < \alpha \le 1$.
Yu~{\it et al.}~\cite{YDLT18, YDCL20} recently improved the ratio to
$\frac{1-1/e}{8(\lceil \frac{4}{\sqrt{3}}\alpha \rceil+1)^2}$.
It can be seen that both the approximation ratios in~\cite{HLS15}
and~\cite{YDLT18, YDCL20} are $\frac{1-1/e}{8(3+1)^2}=\frac{1-1/e}{128}$, when $\alpha=\frac{R_s}{R_c}=1$,
which indicates that the performance of the solutions delivered by
the both algorithms
may be far from the optimal solution.
Therefore, the both algorithms in~\cite{HLS15}
and~\cite{YDLT18, YDCL20} are applicable to the case with many to-be-placed sensors,
i.e., the value of $K$ is very large, e.g., $K=10,000$.
Notice that there are usually tens or hundreds of UAVs to-be-deployed  in a real UAV network, and the approximation ratio $\frac{1-1/e}{\lfloor \sqrt{K} \rfloor}$
in this case is much larger than $\frac{1-1/e}{128}$.
For example, $\frac{1-1/e}{\lfloor \sqrt{K} \rfloor} = \frac{1-1/e}{10} > \frac{1-1/e}{32}$,
when $K (= 100)$ UAVs are to be deployed.

Although the proposed $\frac{1-1/e}{\lfloor \sqrt{K} \rfloor}$-approximation algorithm in this paper
are motivated by the $\frac{1-1/e}{ 5(\sqrt{K} +1)}$-approximation algorithm in~\cite{KLT15},
 the main technical differences between them are twofold.
 (i) One difference is that the algorithm in~\cite{KLT15} shows that a tree $T^*$ that covers the optimal $K$ location nodes can be decomposed into $5(\sqrt{K} +1)$ subtrees  $T_1, T_2, \ldots, T_{5(\sqrt{K} +1)}$ so that the number of nodes in each subtree is no more than $\lfloor \sqrt{K} \rfloor$. Then, there must be one subtree, say $T_l$, among the $5(\sqrt{K} +1)$ subtrees such that the sum of the data rates of users served by UAVs deployed at the location nodes in the subtree is no less than $\frac{1}{5(\sqrt{K} +1)}$ of the sum of  the data rates for the optimal tree $T^*$.  In contrast, in this paper we show that the tree $T^*$  can be decomposed into $\lfloor \sqrt{K} \rfloor$ subtrees $T^{\#}_1, T^{\#}_2, \ldots, T^{\#}_{\lfloor \sqrt{K} \rfloor}$, such that the number of nodes in each subtree is no greater than $2\lfloor \sqrt{K} \rfloor$, see Lemma~\ref{BoundDelta} in Section~\ref{secBoundNumTrees}. It thus can be seen that there must be one subtree, say $T^{\#}_l$, among the $\lfloor \sqrt{K} \rfloor$ subtrees,  such that the sum of the data rates of users served by UAVs at the location nodes in $T^{\#}_l$ is no less than $\frac{1}{\lfloor \sqrt{K} \rfloor}$ of the sum of the data rates for the optimal tree $T^*$. It must be mentioned that the number $2\lfloor \sqrt{K} \rfloor$ of nodes in $T^{\#}_l$  is much larger than the number $\lfloor \sqrt{K} \rfloor$ of nodes in $T_l$.

 (ii) The other difference is that the algorithm in~\cite{KLT15} finds a $(1-1/e)$-approximate tree $T'_l$ of $T_l$ with $\lfloor \sqrt{K} \rfloor$ nodes, by observing that, for any node $v$ in  $T_l$, the minimum number of hops between any node $u$ in $T_l$ and $v$ is no greater than $\lfloor \sqrt{K} \rfloor-1$,
where $e$ is the base of the natural logarithm. Then, $T'_l$ is a $\frac{1-1/e}{5(\sqrt{K} +1)}$-approximate solution.  Contrarily, we find a $(1-1/e)$-approximate tree $T''_l$ of $T^{\#}_l$ with $2\lfloor \sqrt{K} \rfloor$ nodes, by showing that there is a special node $v'$ in $T^{\#}_l$ such that the sum of the minimum numbers of hops between nodes in $T^{\#}_l\setminus\{v'\}$ and $v'$ is no greater than $K-1$, see Lemma~\ref{feasibilityTest} in Section~\ref{secUpperbound}. Then, $T''_l$ is a $\frac{1-1/e}{\lfloor \sqrt{K} \rfloor}$-approximate solution.

We also note that Cadena {\it et al.}~\cite{CMV20}
proposed a $\frac{1-1/e}{5K^{(d-1)/(2d-1)}}$-approximation algorithm
for the problem of finding $K$ connected nodes in a metric graph,
such that a submodular function is maximized, where $d$ is the doubling dimension
of the graph. The proposed algorithm in this paper exhibits some advantages over the one in~\cite{CMV20} as follows.
First, the approximation ratio $\frac{1-1/e}{\sqrt{K}}$ of the proposed algorithm
is better than the approximation ratio  $\frac{1-1/e}{5K^{(d-1)/(2d-1)}}$ of the algorithm
in practical applications.
In a two dimensional Euclidean space, the doubling dimension $d$ is $\log_2 7=2.8$~\cite{W20}.
Then,  $\frac{d-1}{2d-1}=0.39$ and the approximation ratio
of the algorithm by Cadena {\it et al.}~\cite{CMV20} is
$\frac{1-1/e}{5K^{(d-1)/(2d-1)}}=\frac{1-1/e}{5K^{0.39}}$.
It can be seen that the approximation ratio
$\frac{1-1/e}{\sqrt{K}}$ in this paper is larger than $\frac{1-1/e}{5K^{0.39}}$
when $K \le 5^{\frac{1}{0.11}}=2,257,549$.
In a UAV network, there are only tens of, or hundreds  of to-be-deployed UAVs.
On the other hand, the ratio $\frac{1-1/e}{\sqrt{K}}$  is smaller than $\frac{1-1/e}{5K^{0.39}}$ when $K > 2,257,549$.
However, it is unlikely to deploy more than two million UAVs.

Second, the algorithm of Cadena {\it et al.}~\cite{CMV20} is only applicable to metric graphs,
in which the value of the doubling dimension $d$ is small. For a non-metric graph, the value of $d$ may be very large. In this case, the approximation ratio $\frac{1-1/e}{5K^{(d-1)/(2d-1)}}$
approaches to $\frac{1-1/e}{5K^{1/2}}=\frac{1-1/e}{5\sqrt{K}}$, when $d$ is very large.
On the other hand, the proposed algorithm in this paper is applicable to non-metric graphs and its approximation ratio $\frac{1-1/e}{\sqrt{K}}$ still holds.

\section{Preliminaries} \label{secPrelim}
In this section, we first introduce the system  and channel models, then define the problem.

\vspace{-3mm}
\subsection{System model}
When a disaster (e.g., an earthquake or a flooding) occurs, the communication and transportation infrastructures may have been destroyed. To rescue the people trapped in the disaster area, it is urgent to have temporarily emergent communications to help them get out from there. A promising solution is to deploy  multiple UAVs to form a network.

Fig.~\ref{figUAVnetwork} shows  a UAV network in which four UAVs work as base stations to provide communication services (e.g., LTE or WiFi) to affected people in a disaster zone. Assume that at least one of the UAVs serves as a {\it gateway UAV}, which is connected to the Internet, with the help of an emergency communication vehicle or satellites, see Fig.~\ref{figUAVnetwork}. It can be seen that once a trapped people can communicate with a nearby UAV using his smartphone, the people can send and receive critical voice, video, and data to/from the rescue team, with the help of the UAV network.

The disaster area can be treated as a 3D space with length $L$, width $W$, and height $H$, e.g., $L=W=3$~km and $H=500$~m. Assume that there is a set $U$ of $n$ users $u_1, u_2, \ldots, u_n$ on the ground of the disaster area, i.e., $U=\{u_1, u_2, \ldots, u_n\}$.
  We also assume that each user $u_i$ has a minimum data rate requirement
$b^i_{min}$, e.g., $b^i_{min}=2$~kbps.

Denote by $(x_i, y_i, 0)$ the coordinate of a user $u_i$ with $1 \le i \le n$.
We  assume that the locations of  users are known, which
can be obtained by one of the following methods.

On one hand, user smartphones usually are equipped with
GPS modules, and each user can send his location information
to a UAV within the communication range of his smart phone,
when the UAV flies over the disaster area.
On the other hand, if users do not know their locations,
since most UAVs are equipped with GPS modules,
a few UAVs can fly over the disaster area and estimate
user locations, by first taking photos for users with their on-board cameras
(each photo is tagged with the location information where it was taken),
then inferring user locations by applying an existing target detection method~\cite{HS19, KHM+18}.
In addition, if some users are not in Line-of-Sight (LoS) of UAVs and thus cannot be seen by the UAVs, the users can broadcast a probe request with their wireless communication devices. The UAVs then can estimate the locations of the users by the received radio signal strength index (RSSI) measurements~\cite{ESHA21, SPSD21}.

We consider the deployment of  no more than $K$  UAVs to provide communication services (e.g., LTE or WiFi) to  affected users in a monitoring area. Each UAV $k$ is equipped with a lightweight base station device and can act as an aerial base station with $1\le k \le K$~\cite{CGA+16}.
Due to the constraint on the payload of a UAV,
e.g., the maximum payload a  DJI Matrice M300 RTK UAV is only 2.7~kg~\cite{M300-15},
the computation capacity of the base station device
mounted on the UAV
is very limited~\cite{CYA+18, CGA+16, MSC+18}.
Denote by $C$ the {\it service capacity} of each UAV, which means that a UAV can provide communication services to $C$ users simultaneously, e.g., $C=100$ users.

We assume that all UAVs hover at the same altitude $h$, which is the optimal altitude for the maximum coverage from the sky~\cite{AKL14, ZWWW18}, e.g., $h=300$~m.
It can be seen that there are infinite numbers of potential hovering locations for the UAVs, which however makes their placements  intractable. For the sake of convenience, we here only consider a finite number of potential hovering locations for the UAVs, by dividing their hovering plane at altitude $h$ into equal size squares with side length $\delta$, e.g., $\delta=50$ meters. For the sake of convenience, we assume that both length $L$ and width $W$ are divisible by $\delta$. Thus, the hovering plane of the UAVs can be partitioned into $m=\frac{L}{\delta} \times \frac{W}{\delta}$ grids. We further assume that each UAV  hovers only at the center of a grid but do not allow two or more UAVs to hover at the same grid to avoid collisions~\cite{ZWWW18}. Denote by $v_1, v_2, \ldots, v_m$ the center locations of the $m$ grids. Let $V$ be the set of the $m$ potential hovering locations, i.e., $V=\{v_1, v_2, \ldots, v_m\}$.
Table~\ref{notationTable} lists the notations used in this paper.

\begin{table}[tp]
	\centering
	\caption{Notation Table}
	\label{notationTable}
	\footnotesize
	{\begin{tabular}{|p{2.75cm}|p{5cm}|}
			\hline
          	$U=\{u_1, u_2, \ldots, u_n\}$ & set of $n$ ground users $u_1, u_2, \ldots, u_n$\\ \hline
           $b^i_{min}$  &  minimum data rate of a user $u_i$ \\ \hline
           $(x_i, ~y_i,~ 0)$ & coordinate of user $u_i$ with $1\le i \le n$\\ \hline
          $K$ & number of UAVs\\ \hline
          $C$ & service capacity of each UAV\\ \hline
          $B_w$ & bandwidth of each user\\ \hline
          $L, W, H$ &  length, width, and height of the disaster area\\ \hline
          $h$ & hovering altitude of UAVs\\ \hline
         $\delta$ &  side length of a square \\ \hline
           $m=\frac{L}{\delta}\times\frac{W}{\delta}$ &
           number of squares in the plane at altitude $h$\\ \hline
           $v_1, v_2, \ldots, v_m$ &  center locations of the $m$ squares\\ \hline
          $V=\{v_1, v_2, \ldots, v_m\}$ & set of $m$ potential UAV locations\\ \hline
          $R_{uav}$ & transmission range between two UAVs \\ \hline
           $R_{user}$ & transmission range between a user and a UAV \\ \hline
           $d_{i,j}$ & Euclidean distance between user $u_i$ and the UAV at location $v_j$\\ \hline
           $y_j \in\{0, 1\}$ &  whether a UAV is deployed at location $v_j$\\ \hline
           $x_{ij} \in\{0, 1\}$ &   whether user $u_i$ is served
by a UAV deployed at  location $v_j$\\ \hline
           $r_{i,j}$ & data rate of user $u_i$ from the UAV at $v_j$\\ \hline
          $G=(U \cup V, E)$ & UAV network\\ \hline
           $S (\subseteq V)$ & a set of UAV hovering locations\\ \hline
           $f(S)$ &  maximum sum of the data rates of users served by the UAVs deployed at locations in $S$\\ \hline
          $M$ &   a maximum weighted matching in graph $G$\\ \hline
          $T$ &   a minimum spanning tree in graph $G$\\ \hline
          $w(T)$ &  weighted sum of edges in tree $T$,
          i.e., $w(T)=\sum_{e\in T} w(e)$\\ \hline
	\end{tabular}
}
\vspace{-7mm}
\end{table}

\subsection{Channel models}
We consider both UAV-to-UAV and UAV-to-user channel models as follows. UAV-to-UAV channels are mainly dominated by Line-of-Sight (LoS) links, and can be modelled as the free space path loss~\cite{AKL14}. Denote by $R_{uav}$ the communication range of a UAV, i.e.,  two UAVs can communicate with each other if their Euclidean distance is no greater than $R_{uav}$.

On the other hand, the UAV-to-user channel model is more complicated, which must consider both LoS and NLoS (Non-Line-of-Sight) links~\cite{AKL14, ZWWW18}.
Denote by $PL^{LoS}_{i,j}$ and $PL^{NLoS}_{i,j}$ the
 average pathlosses of LoS and NLoS
 for a user $u_i$
from UAV $j$, respectively.
Following the work in~\cite{AKL14},
we have
{\small
\begin{eqnarray}
PL^{LoS}_{i,j} &=& 20 \log_{10} \frac{4\pi f_c d_{i,j}}{c}   + \eta_{LoS},\\
PL^{NLoS}_{i,j} &=& 20 \log_{10} \frac{4\pi f_c d_{i,j}}{c} + \eta_{NLoS},
\end{eqnarray}
}
where $f_c$ is the radio frequency, $d_{i,j}$ is the Euclidean distance between
user $u_i$ and UAV $j$,
 $c$ is the speed of light,
$\eta_{LoS}$ and $\eta_{NLoS}$ are the average shadow fadings for LoS and NLoS links, respectively, and the value of the pair $(\eta_{LoS}, \eta_{NLoS})$
is (0.1~dB, 21~dB), (1~dB, 20~dB), (1.6~dB, 23~dB), (2.3~dB, 34~dB) for
suburban, urban, dense urban, and highrise urban environments, respectively.

Let $P_t$ (in dB) and $g_t$ (in dB) be the signal transmission power
and antenna gain of each UAV, respectively.
Then, the LoS signal-to-noise ratio (SNR)
and the NLoS SNR  for user $u_i$ from UAV $j$ are
{\small
\begin{eqnarray}
SNR^{LoS}_{i, j} &=& 10^{\frac{P_t + g_t - PL^{LoS}_{i,j} - P_N}{10}},\\
SNR^{NLoS}_{i, j} &=& 10^{\frac{P_t + g_t - PL^{NLoS}_{i,j} - P_N}{10}},
\end{eqnarray}
}
respectively,
where $P_N$ (in dB) is the noise power~\cite{YCX+19}.

Denote by $p^{LoS}_{i,j}$ and $p^{NLoS}_{i,j}$ the probabilities of an LoS link and an NLoS
link between user $u_i$ and UAV $j$, respectively,
where $p^{NLoS}_{i,j}+p^{LoS}_{i,j}=1$, and the value of probability $p^{LoS}_{i,j}$ depends on the UAV altitude $h$
and the  horizontal distance between user $u_i$ and UAV $j$~\cite{AKL14}.
Then, the expected data rate $r_{i,j}$ of user $u_i$ served by UAV $j$ is
{\small
\begin{eqnarray}
r_{i,j} &=& p^{LoS}_{i,j} \cdot B_w \cdot \log_2( 1+ SNR^{LoS}_{i,j})+\nonumber\\
&& ~~         p^{NLoS}_{i,j} \cdot B_w  \cdot \log_2( 1+ SNR^{NLoS}_{i,j})
\end{eqnarray}
}
where $B_w$ is the wireless channel bandwidth~\cite{YCX+19}.

Denote by $R_{user}$ the communication range between a UAV and a user~\cite{AKL14}. Note that $R_{user}$ usually is smaller than $R_{uav}$~\cite{LCTXP18}.

\subsection{Spectrum allocations}

To provide communication service to multiple users at the same time,
we assume that the OFDMA technique is used by the UAV.
Denote by $B_{UAV}$ the  spectrum segment
available for each UAV.
Following the LTE standard~\cite{LTE4G}, the spectrum segment used by a base station is
one value among the  1.4, 3, 5, 10, 15, and 20 MHz.
In this paper, we adopt that $B_{UAV}=20$~MHz~\cite{LTE4G}.
To ensure fair sharing of the communication bandwidth among the users, assume that each user uses the {\it same
amount of bandwidth} $B_w$, e.g., $B_w =180$~kHz~\cite{LTE4G},
and at most $C=100$ users can access the UAV at the same time,
since an resource block is the minimum unit of transmission and
is 180 kHz wide~\cite{LTE4G} (notice that $20~MHz-180~kHz \times 100= 2~MHz$
in the total 20~MHz bandwidth cannot be used).

Since some users may be within the transmission ranges of multiple UAVs,
to reduce the interference of such users,
two adjacent UAVs can be allocated with different spectrum segments~\cite{LTE4G}.
On the other hand, to reuse the frequency as efficiently
as possible,   the same spectrum segment
can be reused by two UAVs if they are far away from each other.
Note that the problem of allocating the minimum number of  spectrum segments
is equivalent to the vertex coloring problem in graphs~\cite{B94}.
Specifically, a graph $G_{spectrum}=(S, E_s)$ is first constructed,
where $S$ the hovering locations of the $K$ UAVs,
and there is an edge $(v_i, v_j)$ in $E_s$ between two locations $v_i$ and $v_j$
if their Euclidean distance $d_{i,j}$
is no more than twice the communication range $R_{user}$ of a ground user,
i.e., $d_{i,j} \le 2 R_{user}$. The vertex coloring problem in $G_{spectrum}$ is
to use the minimum number of colors to color vertices in the graph, such that
no two adjacent vertices are colored with the same color.
The vertex coloring problem however is NP-hard,
and the  algorithm in~\cite{B94} can be applied to find an approximate solution to the problem
in polynomial time.

\vspace{-3mm}
\subsection{Problem definition}
We use an undirected graph $G=(U\cup V, E)$ to represent the UAV network, where $U$ is the set of users on the ground of the monitoring area, $V$ is the set of potential hovering locations of UAVs at altitude $h$. There is an edge $(u_i, v_j)$ in $E$ between a user $u_i$ and a hovering location $v_j$ if the Euclidean distance between them  is no greater than $R_{user}$, and there is an edge $(v_j, v_k)$ in $E$ between two hovering locations $v_j$ and $v_k$ if their distance is no more than $R_{uav}$. Notice that the number of available UAVs just after a disaster may be very limited and thus they may not be able to serve all users. In addition, it usually takes time, e.g., one or two days, to purchase new UAVs and install base station devices for rescuing. However, it is very urgent to provide communication services to users. Then, an important problem is to serve as many users as possible  by the available UAVs.

In this paper we consider the {\it connected maximum throughput problem} in $G$,
which is to choose  no more than $K$ hovering locations among all potential hovering locations in $V$ for placing the   UAVs,
 such that  the network throughput, i.e., the sum of the data rates of users served by the
 deployed UAVs, is maximized, subject to that (i) each user $u_i \in U$ can be served by  at most one UAV  within its communication range $R_{user}$;
  (ii) the data rate of user $u_i$ is no less than its minimum data rate $b^i_{min}$ if user $u_i$ is served by a UAV;
 (iii) the number of users served by each UAV is no more than its service capacity $C$; and (iv) the communication network induced by the deployed  UAVs is connected.

The connected maximum throughput problem can also be formulated by an Integer Programming as follows.

We use a binary variable $x_{ij}$ to indicate whether user $u_i$ is associated with a UAV at hovering location $v_j$, i.e., $x_{ij}=1$ if $u_i$ is associated with $v_j$; otherwise, $x_{ij}=0$.  We use another binary variable $y_j$ to indicate whether there is a UAV deployed at location $v_j$, i.e., $y_j=1$ if  a UAV is deployed at location at $v_j$; otherwise, $y_j=0$. The  problem then is to
\begin{equation}
\max_{ x_{i,j}, y_j}~ \sum^n_{i=1} \sum^m_{j=1} x_{ij}\cdot r_{i,j},
\end{equation}
subject to the following constraints.
{\small
\begin{eqnarray}
\sum^n_{i=1} x_{ij} \le C \cdot y_j,\quad~~~~~\forall v_j \in V \label{conCapUAV}\qquad\qquad\qquad\qquad~~~~\\
\sum_{(v_j, v_k)\in \sigma(S)} y_j \cdot y_k \ge 1,~~\text{if $1 < \sum_{v_j \in S} y_j< K-1$},~\forall S \subset V \label{conConnected}\\
 \sum_{v_j \in V} y_j \le K,   \label{conKuavs}\qquad\qquad\qquad\qquad\qquad\qquad\qquad\qquad\quad\\
 \sum^m_{j=1}   (r_{i,j}-b^i_{min}) x_{i,j} ~\ge~0, ~~\forall u_i \in U \qquad\qquad\qquad \quad \label{conMinDataRate}\\
\sum_{v_j \in V} x_{ij} \le 1,~~\forall u_i \in U \label{conUserAssociation}\qquad\qquad\qquad\qquad\qquad\qquad~~~\\
x_{ij}=0,~~\text{if $d_{ij}>R_{user}$},~~~~\forall u_i \in U,~\forall v_j \in V \label{conValidAssociation}\qquad\quad~\\
x_{ij} \in \{0,1\},~~y_j \in \{0,1\},~~\forall u_i \in U,~~\forall v_j \in V,\qquad~~
\end{eqnarray}
}
where Constraint~(\ref{conCapUAV}) ensures that the number of users served by each UAV is no more than its service capacity $C$. Constraint~(\ref{conConnected}) indicates that the UAV communication network is connected. That is, for any proper subset $S$ of $V$, if the number of deployed UAVs in $S$ is between 1 and $K-1$ (i.e., $1 < \sum_{v_j \in S} y_j< K-1$), then there  must be a UAV deployed at a location $v_j$ in $S$ (i.e., $y_j=1$) and another UAV at a location $v_k$ in $V\setminus S$ (i.e., $y_k=1$), such that they can communicate with each other, where $\sigma(S)$ is the set of edges having exactly one endpoint in $S$. Constraint~(\ref{conKuavs}) implies that no more than $K$ UAVs will be deployed.
Constraint~(\ref{conMinDataRate}) shows that the data rate of each user $u_i$ is no less than
its required minimum data rate $b^i_{min}$ if it is served.
Constraint~(\ref{conUserAssociation}) ensures that each user $u_i$ can be served by no more than one UAV. Constraint~(\ref{conValidAssociation}) indicates that user $u_i$ cannot be served by any UAV at location $v_j$ outside of its communication range $R_{user}$, and $d_{ij}$ is the Euclidean distance between $u_i$ and $v_j$.

To deal with the defined connected maximum throughput problem, we here define another problem:  {\it the maximum assignment problem}, which will serve as an important subroutine for the original problem. Given a subset $S \subseteq V$ of hovering locations with $|S|\le K$ such that a UAV has already been deployed at each location in $S$, the problem is how to assign users in $U$ to the UAVs at locations in $S$ so that the sum of the data rates of users  is maximized,  subject to the service capacity $C$ on each UAV
and the minimum data rate $b^i_{min}$ of each user $u_i$.
Notice that the subnetwork induced by the $|S|$ UAVs of the maximum assignment problem may not necessarily be connected. For any subset $S\subseteq V$, denote by $f(S)$ the
 maximum sum of the data rates of users served by UAVs deployed at locations in $S$.

\vspace{-3mm}
\subsection{Submodular functions} \label{secNotions}
Let $V$ be a set of finite elements and $f$ a function with $f: 2^{V} \mapsto \mathbb{R}^{\ge 0}$. For any two subsets $A$ and $B$ of $V$ with $A\subseteq B$ and any element $v \in V\setminus B$, $f$ is {\it submodular} if $f(A\cup \{v\})-f(A)
\ge f(B\cup \{v\}) - f(B)$~\cite{S04}, and $f$ is {\it monotone submodular} if $f(A) \le f(B)$.



\section{Optimal Algorithm for the maximum assignment problem} \label{secMaxAssign}
In this section, we propose an exact algorithm for the maximum assignment problem, which calculates the maximum sum $f(S)$ of data rates of users served by UAVs at locations in a given subset $S \subseteq V$, assuming that a UAV has already been deployed at each location in $S$. This algorithm will serve as a subroutine for the connected maximum throughput problem later. We also show an important property of function $f(S)$, that is, $f(S)$ is nondecreasing and submodular. This property is a cornerstone of the proposed algorithm for the connected maximum throughput problem later.

\vspace{-1mm}
\subsection{An optimal algorithm for calculating $f(S)$}
The basic idea behind the algorithm is to reduce the problem to
the maximum weighted matching problem in an auxiliary bipartite graph,
and an optimal matching to the problem in the auxiliary graph in turn returns an optimal solution to the maximum assignment problem.

\begin{figure}[htp]
\begin{center}
\includegraphics[scale=0.28]{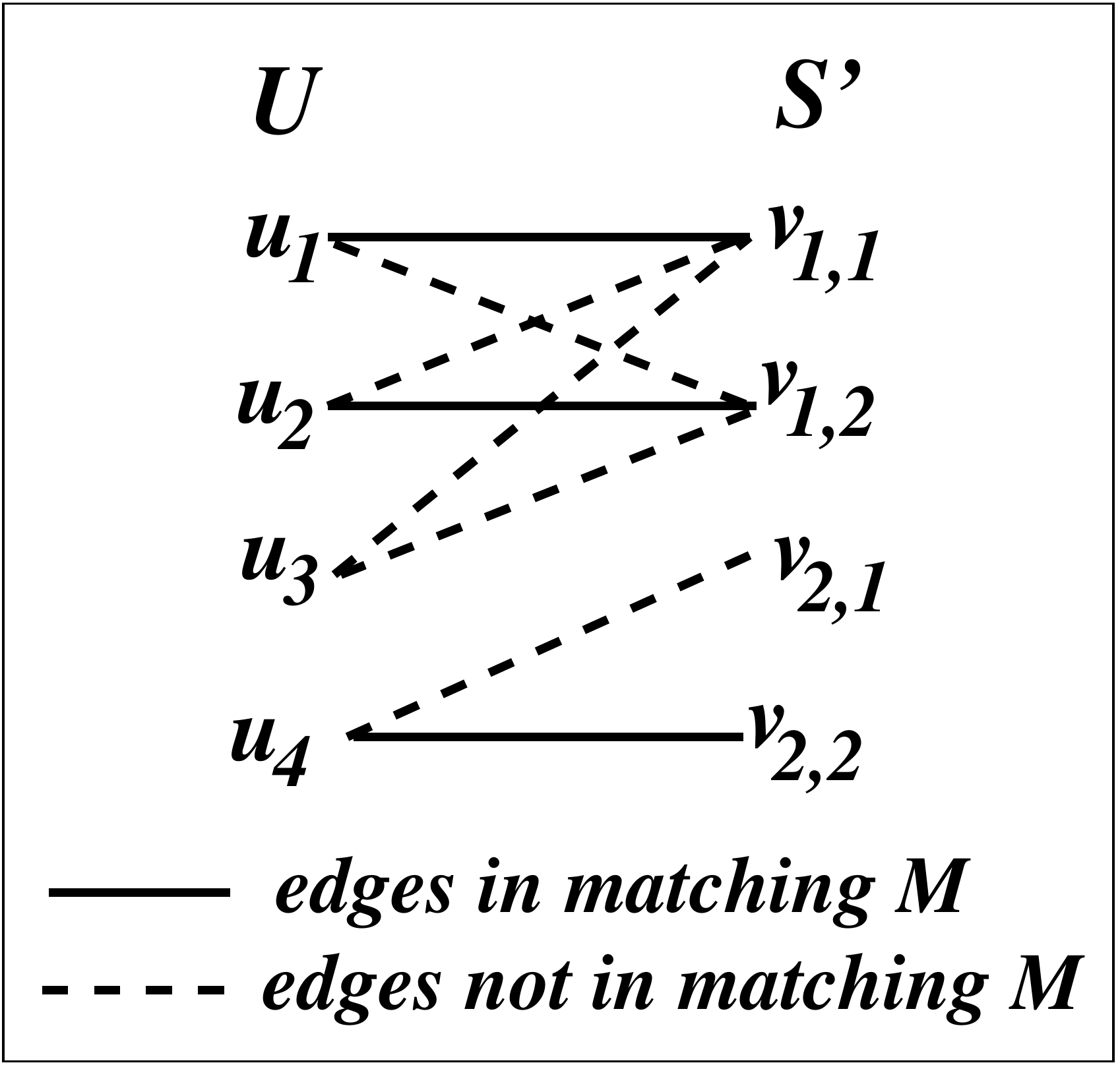}
\vspace{-2mm}
\caption{\small An illustration of the construction of graph $G_S$,
where the service capacity of each UAV is $C=2$,
and users $u_1, u_2,$ and $u_4$ are matched in the maximum weighted matching $M$.}
\label{figBipartiteGraph}
\end{center}
\vspace{-5mm}
\end{figure}

An auxiliary bipartite graph $G_S=(U\cup S', E_S; \rho: E_S \mapsto \mathbb{R}^{\ge 0} )$ is constructed,
where $U$ is the set of users, there are $C$ `{\it virtual}' hovering locations
$v_{j,1}, v_{j,2}, \ldots, v_{j,C}$ in $S'$ for each real location $v_j\in S$, and $C$ is the service capacity of each UAV.
There is an edge $(u_i, v_{j,l})$ in $E_S$ between a user $u_i$
and each virtual location $v_{j,l}$ with $1\le l \le C$ if the
Euclidean distance between user $u_i$ and  $v_j$ is no more than the communication range $R_{user}$ of user $u_i$, i.e., $d_{i,j} \le R_{user}$,
and the data rate $r_{i,j}$ is no less than the minimum data
rate $b^i_{min}$ of user $u_i$, i.e., $r_{i,j} \ge b^i_{min}$.
Finally, the weight $\rho(u_i, v_{j,l})$ of edge $(u_i, v_{j,l})$ is
the data rate $r_{i,j}$ of user $u_i$ if $u_i$ is served by a UAV at $v_j$.
 Fig.~\ref{figBipartiteGraph} illustrates such a graph $G_S$ with $C=2$.

Having constructed graph $G_S$, a maximum weighted matching $M$ in $G_S$ then is  found, by applying an algorithm in~\cite{FT87}, where $M$ is a matching in $G_S$ such
that the weighted sum of edges in $M$, i.e., $\sum_{e\in M} \rho(e)$, is maximized.
Fig.~\ref{figBipartiteGraph} shows  such a maximum weighted  matching $M$. A solution to the  maximum assignment problem is obtained from the matching, where a user $u_i$ is assigned to the UAV at $v_j$ if $u_i$ is matched to a virtual node
$v_{j,l}$ in $M$.  For example, it can be seen  from Fig.~\ref{figBipartiteGraph} that both users $u_1$ and $u_2$ are assigned to $v_1$, and $u_4$ is assigned to $v_2$. However, $u_3$ is not assigned to any one. Then, the sum of the data rates of users served by UAVs is equal to the weighted sum  $\rho(M)$ of edges in matching $M$,
i.e., $\rho(M)=\sum_{e\in M} \rho(e)$.

The algorithm for the maximum assignment problem is presented in {\tt Algorithm}~\ref{AlgMaxAssignment}.

\begin {algorithm}[htp] \small
	  \caption{\small Algorithm for the  maximum assignment problem}
\label{AlgMaxAssignment}
	  \begin{algorithmic}[1]
	  	\REQUIRE  a set $U$ of users, a set $S$ of hovering locations
with a UAV at each location, and the service capacity $C$ of each UAV

	  	\ENSURE An optimal solution to the maximum assignment problem
	  	\STATE Construct an auxiliary bipartite graph $G_S=(U\cup S', E_S; \rho: E_S \mapsto \mathbb{R}^{\ge 0})$,
where  there is an edge $(u_i, v_{j,l})$ in $E_S$ between a user $u_i$
and each virtual location $v_{j,l}$ with $1\le l \le C$, if the
Euclidean distance between user $u_i$ and $v_j$ is no more than the communication range $R_{user}$ of user $u_i$ and  the data rate $r_{i,j}$ is no less than the minimum
data rate $b^i_{min}$ of user $u_i$,
 and $\rho(u_i, v_{j,l})$ is the data rate $r_{i,j}$
of user $u_i$ by the UAV at $v_j$;
        \STATE Find an optimal maximum weighted matching $M$ in $G_S$, by
        invoking the algorithm in~\cite{FT87};
        \STATE For each user $u_i \in U$, assign it to the UAV at $v_j$
        if $u_i$ is matched to a virtual node $v_{j,l}$ in $M$.
	  \end{algorithmic}
  \end{algorithm}

\vspace{-1mm}
\subsection{Algorithm analysis}
\begin{lemma} \label{calfuncF}
Given a UAV network $G=(U\cup V, E)$, a subset $S \subset V$ of hovering locations
with a UAV deployed at each location in $S$,
and  the service capacity $C$ of each UAV,
there is an algorithm, {\tt Algorithm}~\ref{AlgMaxAssignment}, for the maximum assignment problem in $G$, which delivers an optimal solution in time $O((KC+n)^2 \log (KC+n) )$, where $K=|S|$ and $n=|U|$.
\end{lemma}
\begin{IEEEproof}
It can be seen that the value of the optimal solution to the maximum assignment problem
in $G$ is equal to the weighted sum of the edges of the maximum weighted matching $M$ in $G_S$~\cite{FT87}.
Since the algorithm in~\cite{FT87} delivers a maximum weighted matching in $G_S$,
{\tt Algorithm}~\ref{AlgMaxAssignment} delivers an optimal solution.

We analyze the time complexity of {\tt Algorithm}~\ref{AlgMaxAssignment}.
Denote by $n_S$ and $m_S$ the number of nodes and edges in $G_S$, respectively.
Following the construction of $G_S$, it can be seen
that $n_S = KC+n$ and $m_S=O(Cn)$, since the number of UAVs within the communication
range of each user is limited.
 Since the time complexity of the algorithm in~\cite{FT87}
is $O(n^2_S \log n_S + n_S m_S)$, the time complexity of {\tt Algorithm}~\ref{AlgMaxAssignment}
is $O( (KC+n)^2 \log (KC+n) + (KC+n)O(Cn))=O((KC+n)^2 \log (KC+n) )$,
since $C=O(\log (KC+n))$.
The lemma then follows.
\end{IEEEproof}

\begin{figure*}[tp]
\center
\subfigure[nodes in set $V'_j$, where the sum of shortest distances between
nodes in $\{v1, v_2\}$ and $v_j$ is $d_{1,j}+d_{2,j}=2+2=4=K-1$]
{\includegraphics[scale=0.227]{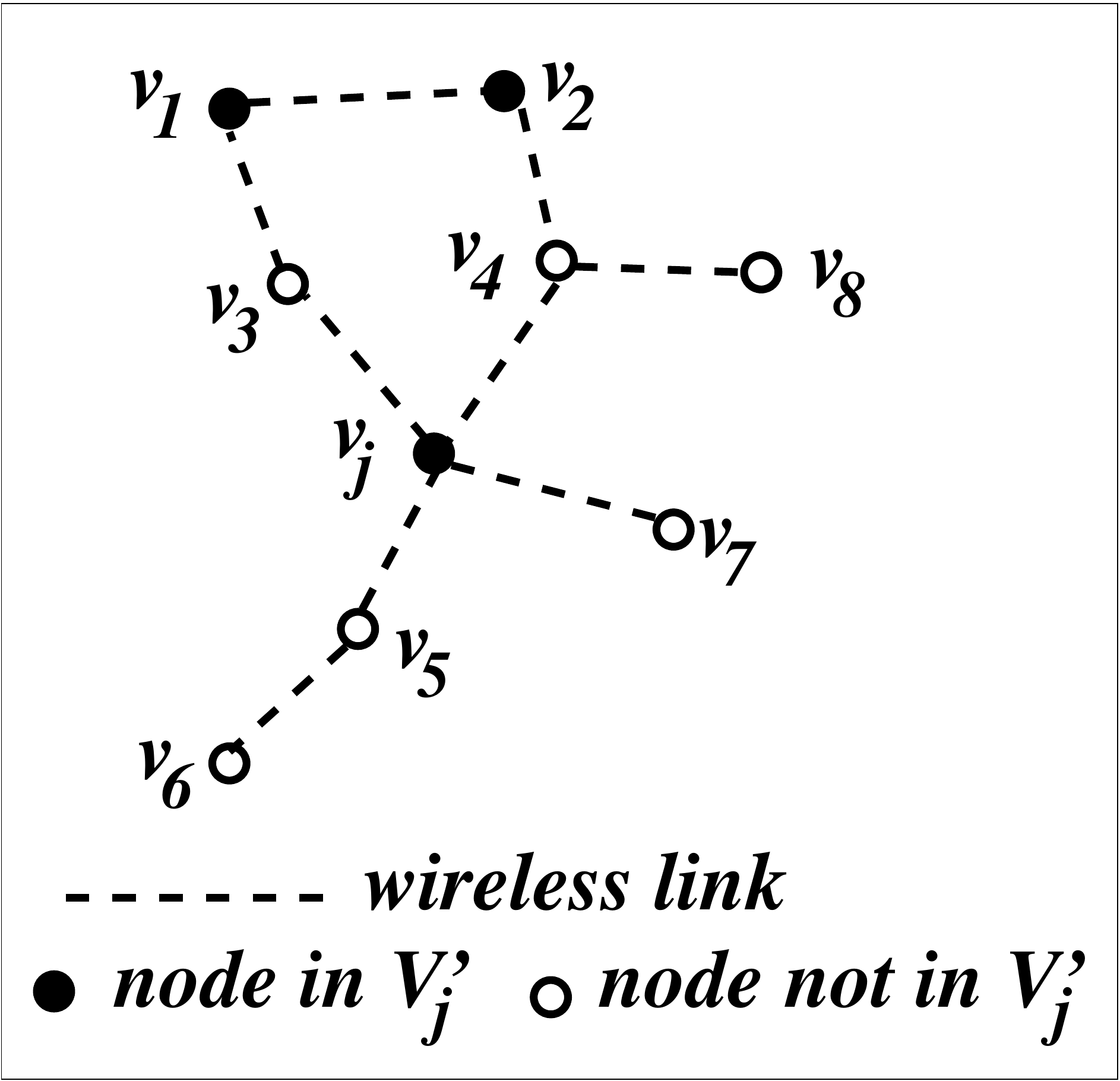}}\quad
\subfigure[graph $G'_j=(V'_j, E'_j)$ and the minimum spanning tree $T'_j$ in $G'_j$]
{\includegraphics[scale=0.23]{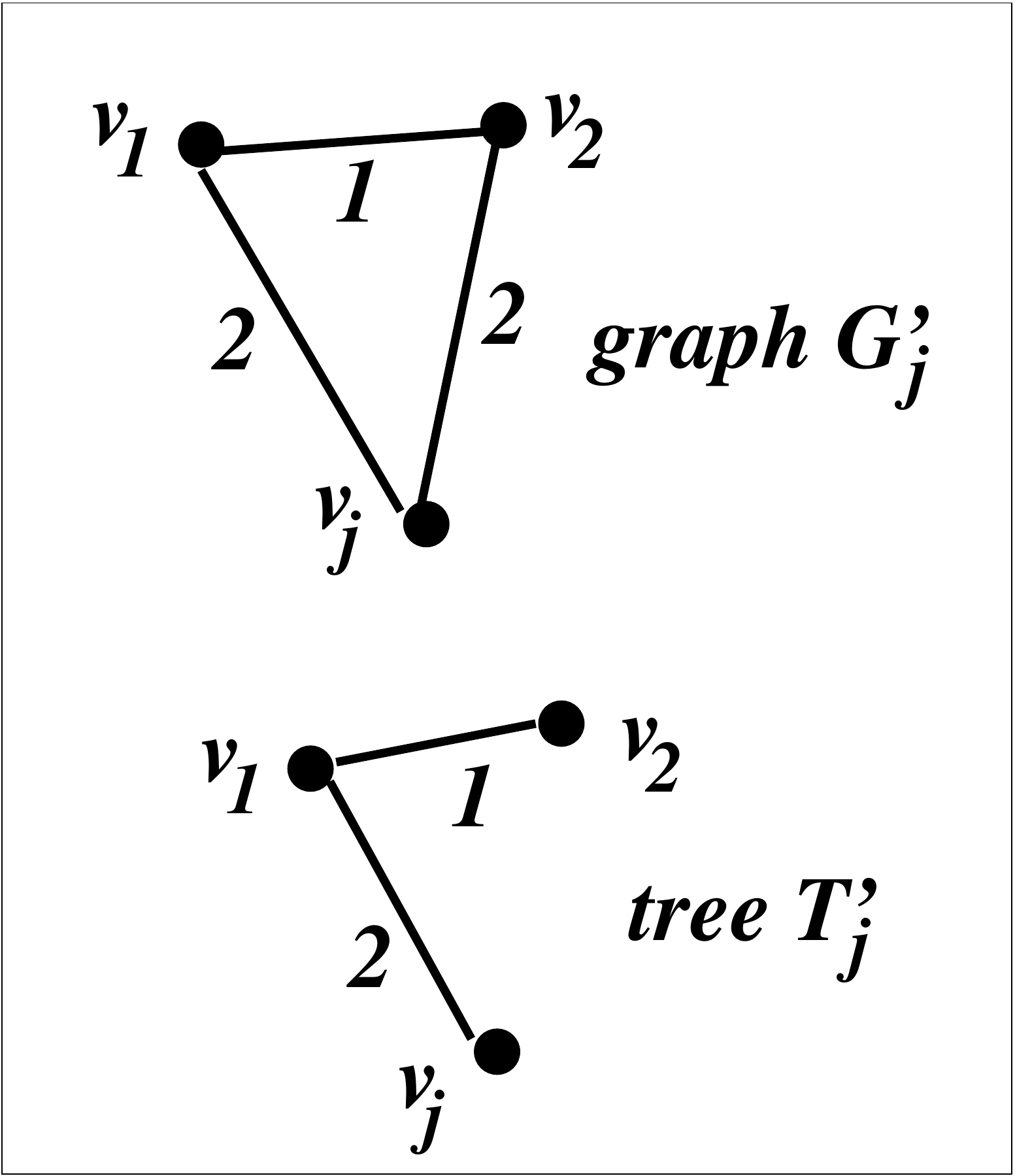}}\quad
\subfigure[graph $G_j$, where $S_j$ is   the set of nodes in $G_j$]
{\includegraphics[scale=0.225]{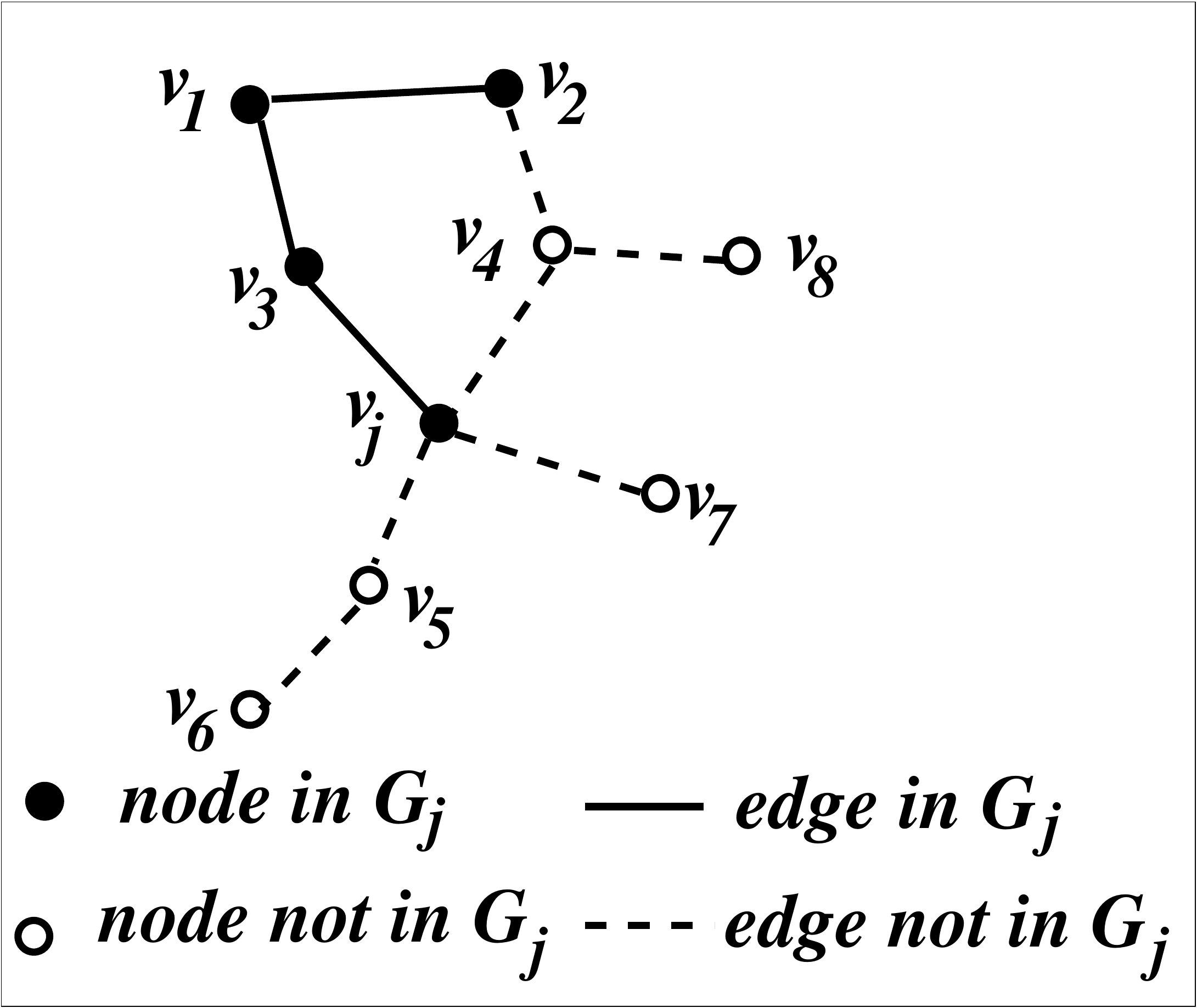}}\quad
\subfigure[nodes in set $S'_j$, which is obtained by adding node $v_5$ to $S_j$]
{\includegraphics[scale=0.234]{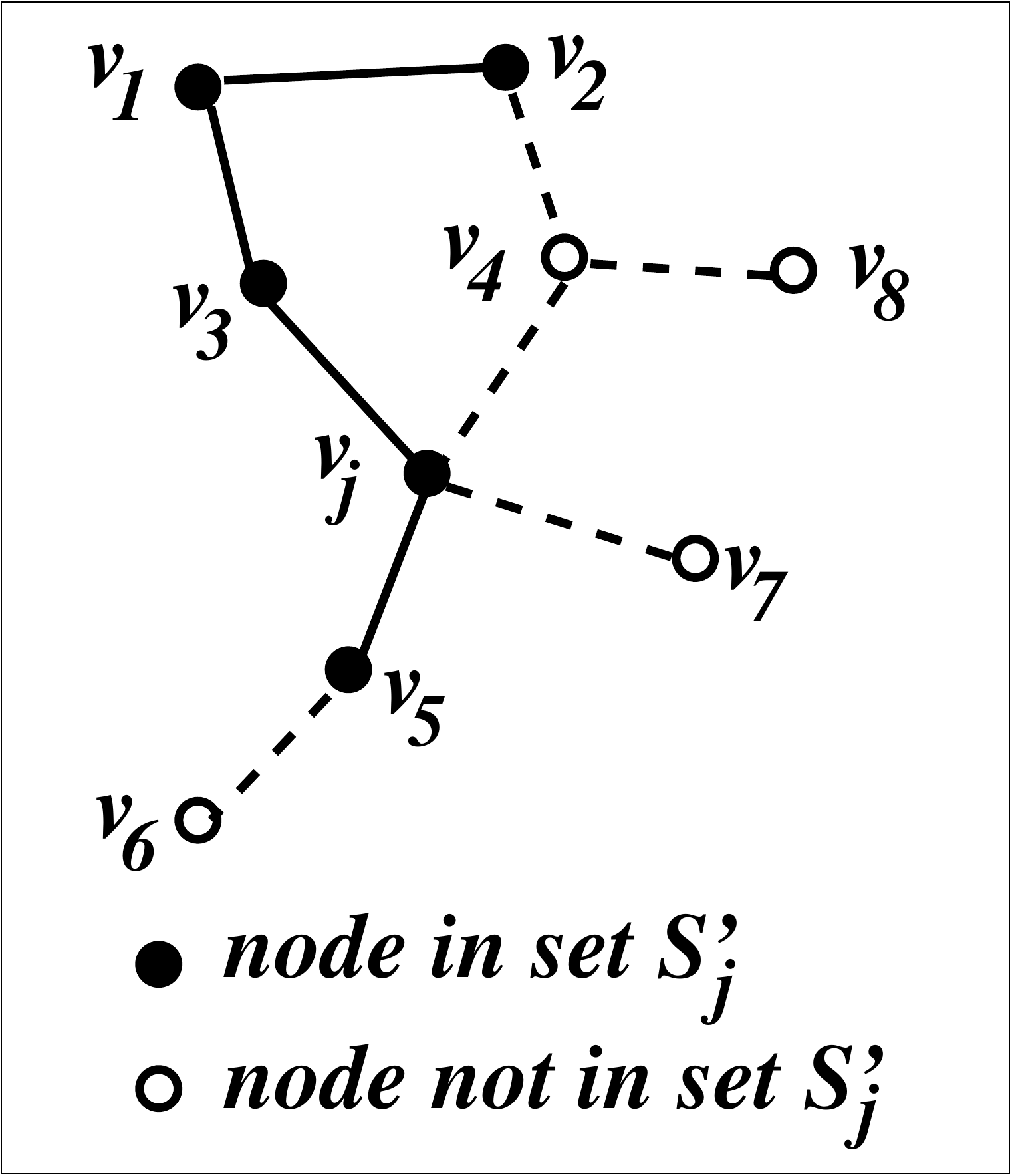}}
\vspace{-4mm}
\caption{An illustration of the execution of the approximation algorithm, where $K=5$.}
	\label{figApproAlg}
\vspace{-6mm}
\end{figure*}

\subsection{Submodularity of function $f(S)$} \label{secSubmodular}

\begin{lemma} \label{lemmaSubmodular}
Given any subset $S$ of $V$, let $f(S)$ be the maximum sum  of the data rates of users served by the UAVs at the hovering locations in $S$, which can be calculated by {\tt Algorithm}~\ref{AlgMaxAssignment}. Then, function $f(S)$ is nondecreasing and submodular.
\end{lemma}
\begin{IEEEproof}
The proof is contained in the supplementary file.
\end{IEEEproof}

\vspace{-3mm}
\section{Approximation Algorithm for the connected maximum throughput problem} \label{secApprAlg}
In this section, we study  the connected maximum throughput problem.
We first provide the basic idea of the proposed approximation algorithm. We then devise  a $\frac{1-1/e}{\lfloor \sqrt{K} \rfloor}$-approximation algorithm for the problem,
where $e$ is the base of the natural logarithm
and $K$ is the number of UAVs.

\vspace{-3mm}
\subsection{Basic idea}
The basic idea behind the proposed algorithm is as follows.
Given any location $v_j \in V$, the algorithm first identifies a subset $V_j$  of hovering locations
around $v_j$,
such that the sum of the data rates of users served by the UAVs at the identified locations
is maximized, while ensuring that the sum of shortest distances between $v_j$ and nodes in $V_j$ is no more than $K-1$, where the shortest distance between $v_l$ and a node $v_k \in V_j$ is the minimum number of hops between them in $G$. Since the subnetwork induced by nodes in $V_j \cup \{v_j\}$ may not be connected,  ensuring these nodes to be connected is done through adding relaying nodes among them while keeping
 the number of nodes in the resulting connected component is no greater than $K$.

\vspace{-3mm}
\subsection{Approximation algorithm}
For each hovering location $v_j\in V$,
the algorithm finds a set $S'_j$ of $K$ location nodes
such that the induced graph $G[S'_j]$  of $G$ by the nodes in $S'_j$ is connected, where $v_j$ is contained in $S'_j$. The solution to the connected maximum throughput problem then is such a set $S'_{j^*}$ that the sum of the data rates of users served by the UAVs deployed at locations in $S'_{j^*}$ is maximized, i.e., $j^*=\arg\max_{1 \le j \le m} \{f(S'_j)\}$, where $m$ is the number of hovering locations in $V$. In the following, we show how to find  set $S'_j$ for each $v_j\in V$.

For each hovering location $v_j\in V$,
we first find the shortest distance $d_{k,j}$ (in terms of numbers of hops)
in $G$ between $v_j$ and each hovering location
$v_k$ in $V\setminus \{v_j\}$, by applying a Breadth-First-Search starting from $v_j$.


We then consider a {\it constrained maximum throughput problem},
which is to
find a subset $V_j$ of $V\setminus\{v_j\}$
such that the sum of the data rates of users served by the UAVs deployed at the locations
in $V_j \cup \{v_j\}$ is maximized, subject to that
the sum of the shortest distances between the nodes in $V_j$ and $v_j$
is no greater than $K-1$,
i.e., $\sum_{v_k \in V_j} d_{k,j} \le K-1$,
where $d_{k,j}$ is the minimum number of hops between a node $v_k \in V_j$ and $v_j$
in  $G$.
We later  show that this problem
can be cast as a submodular function maximization problem subject to a knapsack constraint.
Then, we can find a $(1-1/e)$-approximate solution $V_j$
to the constrained maximum throughput problem, by
applying the algorithm in~\cite{S04},
where $e$ is the base of the natural logarithm.
Let $V'_j = V_j \cup \{v_j\}$, see Fig.~\ref{figApproAlg}(a).

Notice that the induced graph $G[V'_j]$ of $G$ by the nodes in $V'_j$ may not
be connected.
The rest is to find a connected subgraph $G_j$ of $G$,
such  that the nodes in $V'_j$ are contained in $G_j$ and $G_j$ contains no more than $K$ nodes.

A graph $G'_j=(V'_j, E'_j)$ is first constructed from set $V'_j$, where there is an edge
$(v_k,v_l)$ in $E'_j$ between any two nodes $v_k$ and $v_l$ in $V'_j$, and its edge weight $w(v_k,v_l)$
is  the minimum number of hops between them in $G$.
A minimum spanning tree (MST) $T'_j$ in $G'_j$  is then found, see Fig.~\ref{figApproAlg}(b).
There is an important property, that is,
the weighted sum of the edges in $T'_j$ is no greater than $K-1$,
i.e., $w(T'_j)=\sum_{(v_k, v_l) \in T'_j} w(v_k, v_l) \le K-1$,
which will be shown later.
Denote by $n_j$  the number of nodes in tree $T'_j$.
Then, there are $(n_j-1)$ edges in $T'_j$.
For each edge $(v_k, v_l)$ in tree $T'_j$,
there is a corresponding shortest path $P_{k,l}$  in  graph $G$
between nodes $v_k$ and $v_l$.
A {\it connected subgraph} $G_j$ of $G$  then can be obtained from  $T'_j$,
which is the union of the $(n_j-1)$ shortest paths in $G$,
i.e., $G_j=\{P_{k,l}~|~(v_k,v_l) \in T'_j\}$.
For example, Fig.~\ref{figApproAlg}(c) shows  a graph $G_j$
constructed from the tree $T'_j$ in Fig.~\ref{figApproAlg}(b),
where $P_{1,j}=v_1-v_3-v_j$ and $P_{1,2}=v_1-v_2$.
Following the construction of $G_j$, it can be seen that
the number of edges in $G_j$ is no more than the weighted sum $w(T'_j)$ of the edges in $T'_j$, as the weight $w(v_k, v_l)$ of each edge $(v_k, v_l)$ in $T'_j$
is the shortest distance between nodes $v_k$ and $v_l$ in $G$.
The number of nodes in $G_j$ thus is no more than $w(T'_j)+1 \le K-1 +1 = K$,
since $G_j$ is connected.

Denote by $S_j$ the node set of graph $G_j$.
Then, $|S_j| \le K$.
We construct a set $S'_j$ that contains
 no more than $K$ nodes and the nodes in $S_j$  are contained in $S'_j$ as well.
If $|S_j|=K$, then $S'_j=S_j$.
Otherwise ($|S_j| < K$),
let $S'_j=S_j$ initially.
We add nodes to $S'_j$ one by one
until $S'_j$ contains exactly $K$ nodes
or the marginal gain of adding any node is zero,
such that
each added node has the maximum marginal gain,
subject to
that the node is connected with a node already in $S'_j$.
For example, Fig.~\ref{figApproAlg}(d) shows that
node $v_5$ is added to $S'_j$ and $|S'_j|=K=5$.

It can be seen that the induced subgraph $G[S'_j]$ by the nodes in set $S'_j$ is connected.


The algorithm for the problem is presented in
{\tt Algorithm}~\ref{ApproAlg}.

\begin {algorithm}[htp]\small
	  \caption{Approximation algorithm for the  connected maximum throughput problem
({\tt ApproAlg})}
\label{ApproAlg}
	  \begin{algorithmic}[1]
	  	\REQUIRE  a set $U$ of users, a set $V$ of potential hovering locations,
and $K$ UAVs with the service capacity $C$ of each UAV
	  	\ENSURE A solution to the connected maximum throughput problem
        \STATE Let $D$ be the maximum number between $2\lfloor \sqrt{K-1}\rfloor$
        and  the largest odd number no more than $\lfloor\sqrt{4K-3}\rfloor$;
        \STATE Let $V'\leftarrow \emptyset$; /* the  set of hovering locations */
        \FOR{each location $v_j \in V$}
            \STATE Calculate the shortest distance between each  location in $V\setminus \{v_j\}$ and $v_j$ in $G$, by using a Breadth-First-Search
            starting from $v_j$,
            where the cost of each edge is one;
            \STATE Find a set $V_j$ of locations   for the  constrained maximum throughput problem, by invoking the algorithm in~\cite{S04}; \label{SetNotConnected}
            \STATE Let $V'_j = V_j \cup \{v_j\}$;
            \STATE Construct a graph $G'_j=(V'_j, E'_j)$, where
            there is an edge $(v_k, v_l)\in E'_j$ between any two nodes
            $v_k$ and $v_l$ in $V'_j$, and its edge weight $w(v_k, v_l)$
            is the minimum number of hops between $v_k$ and $v_l$ in $G$;
            \STATE Find a Minimum Spanning Tree (MST) $T'_j$ in $G'_j$; \label{mst}
            \STATE Construct a subgraph $G_j$ of $G$,
            where $G_j=\{P_{k,l}~|~(v_k, v_l)\in T'_j\}$
            and $P_{k,l}$ is the shortest path in $G$ between nodes $v_k$ and $v_l$. Let $S_j$ be the set of nodes in $G_j$;\label{setConnected}
            \STATE Let $S'_j=S_j$ initially, and continue to
            add nodes in $V\setminus S'_j$ to $S'_j$ until $S'_j$ has $K$ nodes,
            such that
each added node has the maximum positive marginal gain,
            subject
            to that the node is connected to a node already in $S'_j$;
            \IF{$f(S'_j)> f(V')$}
                \STATE /* Find a better set of hovering locations */
                \STATE Let $V'\leftarrow S'_j$;
            \ENDIF
        \ENDFOR
        \STATE Assign users in $U$ to the UAVs at the hovering locations in $V'$
        by invoking {\tt Algorithm}~\ref{AlgMaxAssignment};\\
        \RETURN the hovering locations in $V'$ and the assignment of users in $U$.
	  \end{algorithmic}
  \end{algorithm}

\vspace{-3mm}
\subsection{Redeployment of UAVs with user mobility}
Users may move around in the disaster area.
It can be seen that an optimal deployment of the UAVs may become sub-optimal  after a period of time, due to users mobility. In this case,
we invoke the proposed algorithm to calculate the updated optimal deployment locations of the $K$ UAVs
 every time slot, e.g., 2 minutes.
At the beginning of each time slot,
we first calculate the new deployment locations
 of the UAVs with the most recent location information of users,
by invoking the proposed algorithm.
taken by the on-board cameras of the UAVs~\cite{HS19, KHM+18}.
If the network throughput under the previous UAV deployment locations
is only slightly worse than that under  this new  UAV deployment
locations, e.g., no more than 5\% smaller, the $K$ UAVs do not fly to their new deployment locations, since frequent redeployments of UAVs
consume large amounts of energy. Otherwise (the previous network throughput
is at least 5\% smaller than  the new network throughput),
the UAVs fly to their new locations.

\subsection{The energy issue of UAVs}

To provide uninterrupted communication services to users in a disaster area
for a critical period, e.g., within 72 hours, we assume that the UAV communication  network  consists of
(1) $K$, e.g., 30, communication UAVs; (2) $K_{sUAV}$ standby communication UAVs; and (3) $K_{battery}$ standby UAV batteries which can
be simultaneously charged at a nearby service center, where
the cost of a UAV battery usually is much cheaper than the cost of a UAV.
When some communication UAVs run out of energy, the standby UAVs
first replace the communication UAVs by flying to the service hovering locations of the communication UAVs, and the communication UAVs then return to the service center to replace their batteries. The communication UAVs act as
new standby UAVs for later UAV replacements.
The detached UAV batteries can be recharged at the service center.
 By doing so, there are always $K$ communication UAVs deployed to provide
communication services to ground users.
The calculations of the number $K_{sUAV}$  of standby communication UAVs
and the number $K_{battery}$  of standby UAV batteries are contained
in the supplementary file.

\vspace{-4mm}
\section{Analysis of the Approximation Algorithm} \label{secAlgAnalysis}
In this section, we analyze the approximation ratio of the proposed algorithm, {\tt Algorithm}~\ref{ApproAlg}, for the connected maximum throughput problem. Denote by $V^*$ and $OPT$ the optimal solution and its value, respectively, i.e., $OPT=f(V^*)$. Following the definition of the  connected maximum throughput problem, the induced graph $G[V^*]$ by the nodes in $V^*$ is connected. It can be seen that there is a  spanning tree $T$ in $G[V^*]$, assuming that the cost of each edge is one. The roadmap of the approximation ratio analysis is as follows.

We first show that tree $T$ can be decomposed into
$\Delta$ subtrees $T_1, T_2, \ldots, T_{\Delta}$
such that the number of nodes in each subtree
is no more than $D$ and $\Delta \le \lfloor \sqrt{K} \rfloor$
in Lemma~\ref{BoundDelta} of Section~\ref{secBoundNumTrees},
where $D$ be the maximum integer between  $2\lfloor \sqrt{K-1}\rfloor$
and the largest odd number no
more than $\lfloor \sqrt{4K-3}\rfloor$.
Due to the submodularity of the objective function by Lemma~\ref{lemmaSubmodular}
in  Section~\ref{secSubmodular}, it can be seen that there is a subtree,
 say $T_l$, among the $\Delta$ subtrees such that
the sum of the data rates of users served by the UAVs in $T_l$ is no less than $\frac{1}{\Delta}$ of that in the original tree $T$,
i.e., $f(T_l) \ge  \frac{f(T)}{\Delta} = \frac{OPT}{\Delta}
\ge \frac{OPT}{\lfloor \sqrt{K} \rfloor}$, where $f(T)= OPT$.

We then prove that there is a node $v_l$ in $T_l$ such that the sum of
the shortest distances between $v_l$ and the nodes in $T_l \setminus \{v_l\}$
is no more than $K-1$, i.e., $\sum_{v_k \in T_l \setminus \{v_l\}} d_{k,l} \le K-1$,
see  Lemma~\ref{feasibilityTest} in Section~\ref{secUpperbound}. Denote by $V^*_l$ the optimal solution to the constrained maximum throughput problem
with respect to node $v_l$.
It can be seen that the nodes in $T_l \setminus \{v_l\}$ form a feasible solution
to the constrained maximum throughput problem with respect to node $v_l$,
and thus  $f(T_l) \le f(V^*_l \cup \{v_l\})$.
Therefore, we  obtain a non-trivial upper bound on the optimal solution $OPT$,
i.e., $OPT \le \lfloor \sqrt{K} \rfloor \cdot f(T_l) \le \lfloor \sqrt{K} \rfloor \cdot f(V^*_l \cup \{v_l\})$, see Lemma~\ref{optUpperBound} in Section~\ref{secUpperbound}.

 On the other hand, we are able to find a $(1-1/e)$-approximate solution $V_l$ to the constrained maximum throughput problem with respect to node $v_l$,
which implies that $f(V_l \cup \{v_l\}) \ge (1-1/e)\cdot f(V^*_l \cup \{v_l\}) \ge (1-1/e)\frac{OPT}{\lfloor \sqrt{K} \rfloor}$. In addition, we can obtain a set $S_l$ of $K$ nodes including nodes
in  $V_l \cup \{v_l\}$ such that the induced subgraph by $S_l$ in $G$ is connected,
since $\sum_{v_k \in V_l \setminus \{v_l\}} d_{k,l} \le K-1$.
Then, $f(S_l) \ge f(V_l \cup \{v_l\}) \ge \frac{1-1/e}{\lfloor \sqrt{K} \rfloor} \cdot OPT$, as $f(.)$ is a nondecreasing function. $S_l$ thus is a $\frac{1-1/e}{\lfloor \sqrt{K} \rfloor}$-approximate solution
to the connected maximum throughput problem, see  Theorem~\ref{theoremApproRatio} in Section~\ref{secRatioAnalysis}.


\subsection{Bound the number of decomposed subtrees} \label{secBoundNumTrees}

It can be seen that there are $K-1$ edges in tree $T$ since $|V^*|=K$.
Let $B=D-1$, where $D$ is the maximum number between $2\lfloor \sqrt{K-1}\rfloor$
and  the largest odd number no more than $\lfloor\sqrt{4K-3}\rfloor$.
Following the work due to Xu {\it et al.}~\cite{WLL15},
tree $T$ can be decomposed into  $\Delta$  edge-disjoint subtrees $T_1, T_2, \ldots, T_{\Delta}$,
such that the number of edges in each subtree is no more than $B$,
where $\Delta ~\le~  \lfloor \frac{K-1}{ (B+1)/2 }\rfloor
~=~ \lfloor \frac{2(K-1)}{ D}\rfloor.$
Then, the number of nodes in each of the $\Delta$ subtrees is no more than
$B+1=D$.

We show that $\Delta \le \lfloor \sqrt{K} \rfloor$ by the following lemma.

\begin{lemma} \label{BoundDelta}
Given a tree $T$ with  $K$ nodes and $K\ge 2$,
let $D$ be the maximum integer between  $2\lfloor \sqrt{K-1}\rfloor$
and the largest odd number no
more than $\lfloor \sqrt{4K-3}\rfloor$.
Then,  tree $T$  can be decomposed into $\Delta$ subtrees
$T_1, T_2, \ldots, T_{\Delta}$ such that the number of nodes in each subtree
is no more than $D$ and $\Delta \le \lfloor \sqrt{K} \rfloor$.
\end{lemma}
\begin{IEEEproof}
The proof is contained in the supplementary file.
\end{IEEEproof}

\subsection{An upper bound on the optimal solution} \label{secUpperbound}
In the following we provide a non-trivial upper bound on the optimal solution to the
connected maximum throughput problem, which will be used in  the approximation ratio analysis.

For each subtree $T_l$ with $1\le l \le \Delta$,
let $P_l$ be the longest path in $T_l$,
where the length of a path in $T_l$ is the number of edges in the path.
Also, let $v_l$ be a middle node of path $P_l$.
For node  $v_l \in V$, denote by $V^*_l$
the optimal solution to the constrained maximum throughput problem.
In the following, we first show that the sum of the shortest distances between nodes
in $V(T_l)\setminus \{v_l\}$ and $v_l$ is no more than $K-1$.
Then, $V(T_l)\setminus \{v_l\}$ is a feasible solution to the constrained maximum throughput problem. Therefore, $f(T_l)=f( (T_l\setminus \{v_l\})\cup \{v_l\})
\le f(V^*_l \cup \{v_l\})$.

\begin{lemma} \label{feasibilityTest}
Given any tree $T_l$ in graph $G$ with no more than $D$ nodes, let $P_l$ be the longest path in $T_l$ and $v_l$ be a middle node of path $P_l$, where $D$ is the maximum integer between  $2\lfloor \sqrt{K-1}\rfloor$
and the largest odd number no
more than $\lfloor \sqrt{4K-3}\rfloor$.
Then,  the sum  of the shortest distances between nodes
in $V(T_l)\setminus \{v_l\}$ and $v_l$ in $G$ is no more than $K-1$,
i.e., $\sum_{v_k \in V(T_l)\setminus \{v_l\}} d_{k,l} \le K-1$.
\end{lemma}
\begin{IEEEproof}
The proof is contained in the supplementary file.
\end{IEEEproof}

\begin{lemma} \label{optUpperBound}
Denote by $V^*$ and $OPT$ the optimal solution and its value of the connected maximum throughput problem, i.e., $OPT=f(V^*)$. For each location  $v_j \in V$, denote by $V^*_j$
the optimal solution to the constrained maximum throughput problem.
Then, $OPT \le \lfloor \sqrt{K} \rfloor \cdot\max_{v_j\in V} \{f(V^*_j \cup \{v_j\})\}$.
\end{lemma}
\begin{IEEEproof}
It can be verified that this claim holds when $K=1$.
In the following, we assume that $K\ge 2$.

 Recall that, by Lemma~\ref{BoundDelta},
 tree $T$ can be decomposed into  $\Delta$  edge-disjoint subtrees $T_1, T_2, \ldots, T_{\Delta}$,
such that the number of nodes in each subtree is no more than $D$
and $\Delta \le  \lfloor \sqrt{K}\rfloor$.
In addition, for each subtree $T_l$ with $1\le l \le \Delta$,
let $P_l$ be the longest path in $T_l$
and $v_l$ be a middle node of path $P_l$.
For node  $v_l \in V$, denote by $V^*_l$
the optimal solution to the constrained maximum throughput problem.
Recall that  the sum of the shortest distances between nodes
in $V(T_l)\setminus \{v_l\}$ and $v_l$ is no more than $K-1$, by Lemma~\ref{feasibilityTest}.
Then, $V(T_l)\setminus \{v_l\}$ is a feasible solution to the constrained maximum throughput problem. Therefore,
\begin{eqnarray}
f(T_l)&=&f( (V(T_l)\setminus \{v_l\})\cup \{v_l\}) \nonumber \\
&\le& f(V^*_l \cup \{v_l\})  ~\le~ \max_{v_j\in V} \{f(V^*_j \cup \{v_j\})\}. \label{upperBound}
\end{eqnarray}


On the other hand, we  have
{\small
\begin{eqnarray}
OPT &=& f(V^*) ~\le~ \sum^\Delta_{l=1} f(T_l),~\text{as $f(.)$ is a submodular} \nonumber\\
    &=& \Delta \cdot \max_{v_j\in V} \{f(V^*_j \cup \{v_j\})\},~\text{by Ineq.~(\ref{upperBound})}  \nonumber\\
    &\le& \lfloor \sqrt{K} \rfloor\cdot \max_{v_j\in V} \{f(V^*_j \cup \{v_j\})\},~\text{by Lemma~\ref{BoundDelta}.}
    \label{BoundKlarger2}
\end{eqnarray}
}
The lemma then follows.
\end{IEEEproof}

\subsection{The approximation ratio analysis} \label{secRatioAnalysis}

We finally analyze the approximation ratio of the proposed algorithm
by the following theorem.
\begin{theorem} \label{theoremApproRatio}
Given a UAV network $G=(U\cup V, E)$ and $K$ UAVs with the service capacity
constraint $C$ on each UAV, there is a $\frac{1-1/e}{\lfloor \sqrt{K} \rfloor}$-approximation algorithm,
{\tt Algorithm}~\ref{ApproAlg},
for the connected maximum throughput problem with a time complexity $O(K m^3 n^2 \log n)$,
where $e$ is the base of the natural logarithm, $n=|U|$, and $m=|V|$.
\end{theorem}
\begin{IEEEproof}
The feasibility of set $V'$ is proved in
 in the supplementary file.
The rest is to analyze its approximation ratio.

\begin{figure*}[ht]
\vspace{-2mm}
\center
\subfigure[Network throughput]
{\includegraphics[scale=0.21]{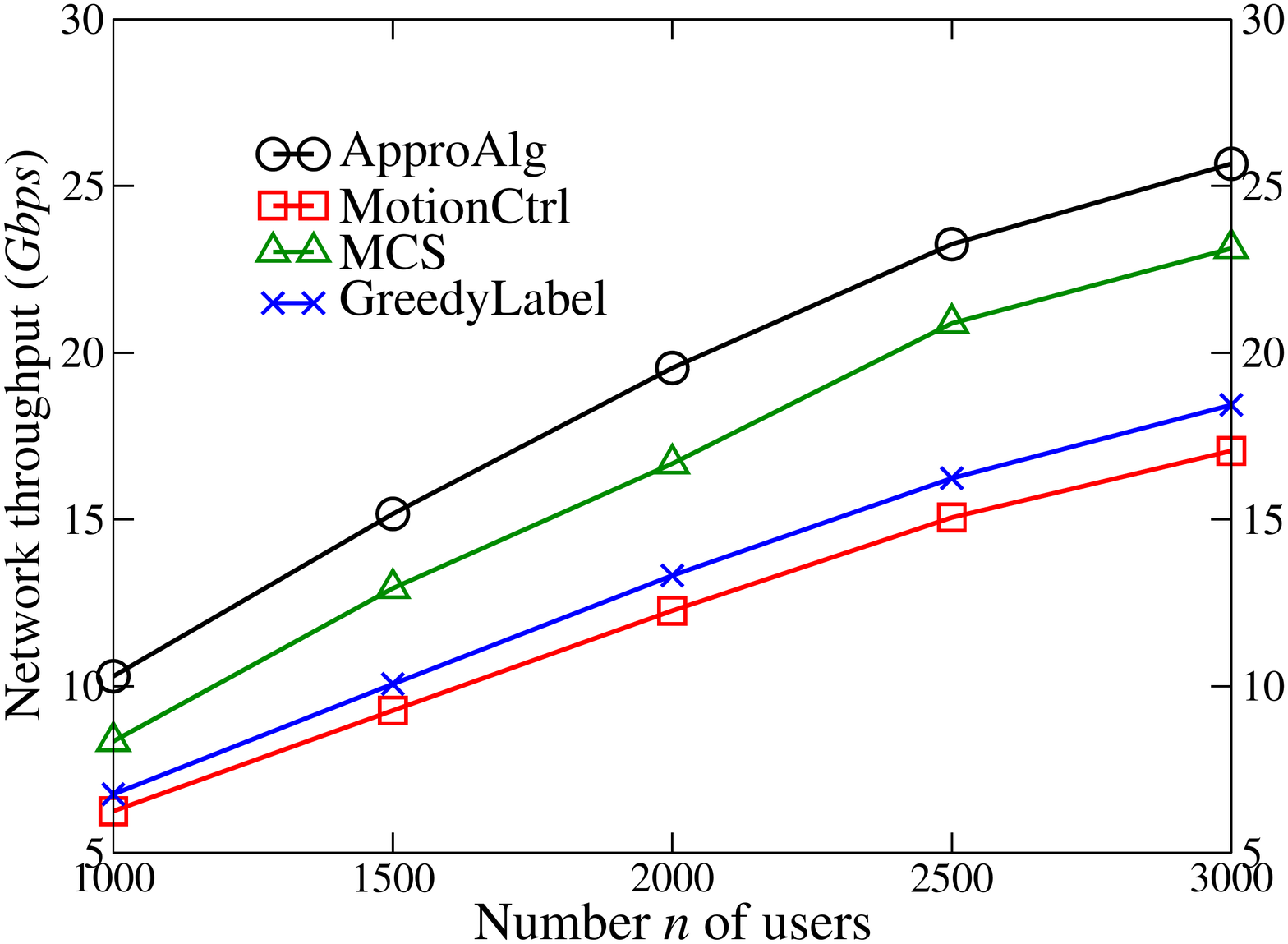}}~
\subfigure[Flying energy consumption per UAV]
{\includegraphics[scale=0.21]{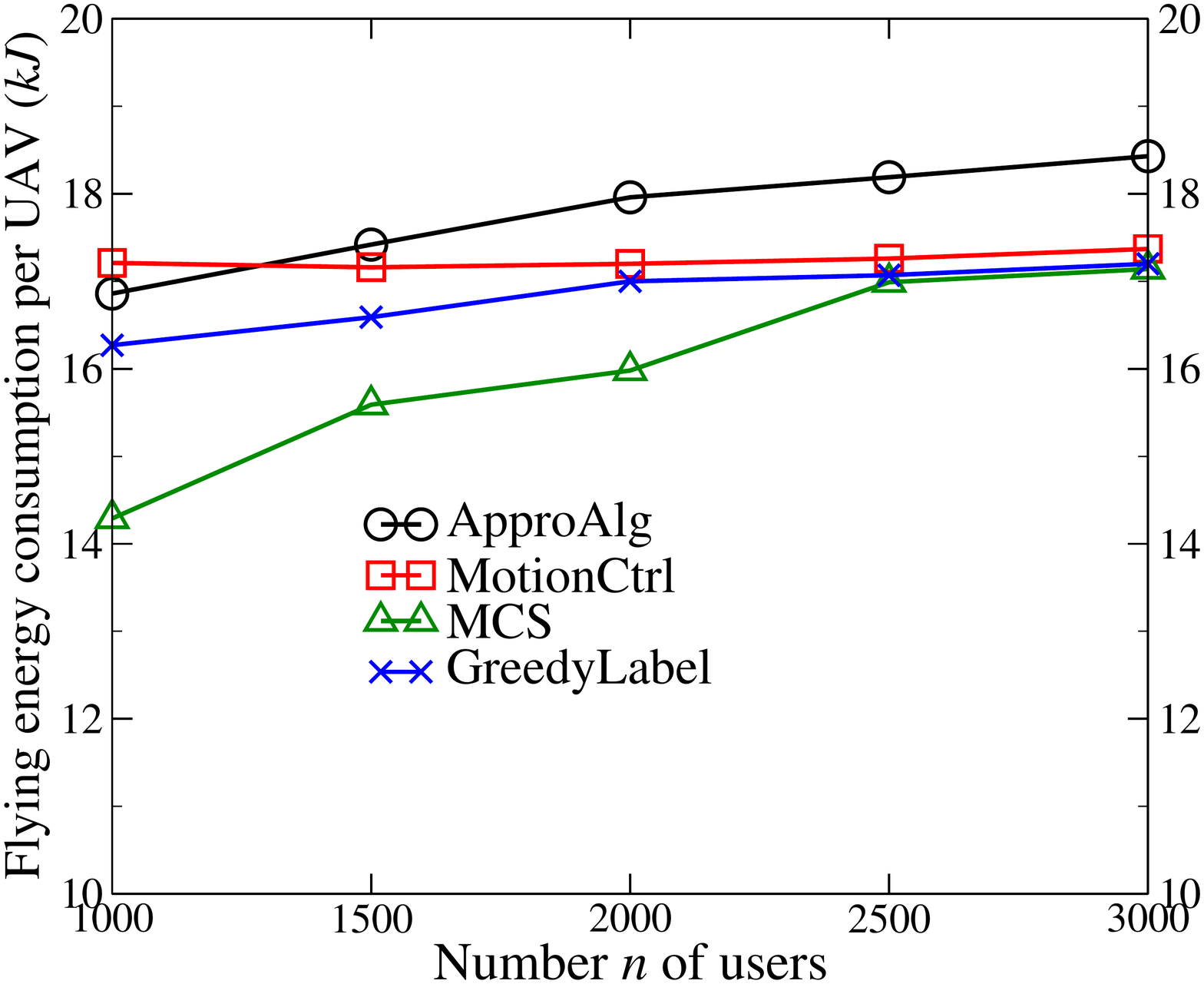}}~
\subfigure[Algorithm running time]
{\includegraphics[scale=0.21]{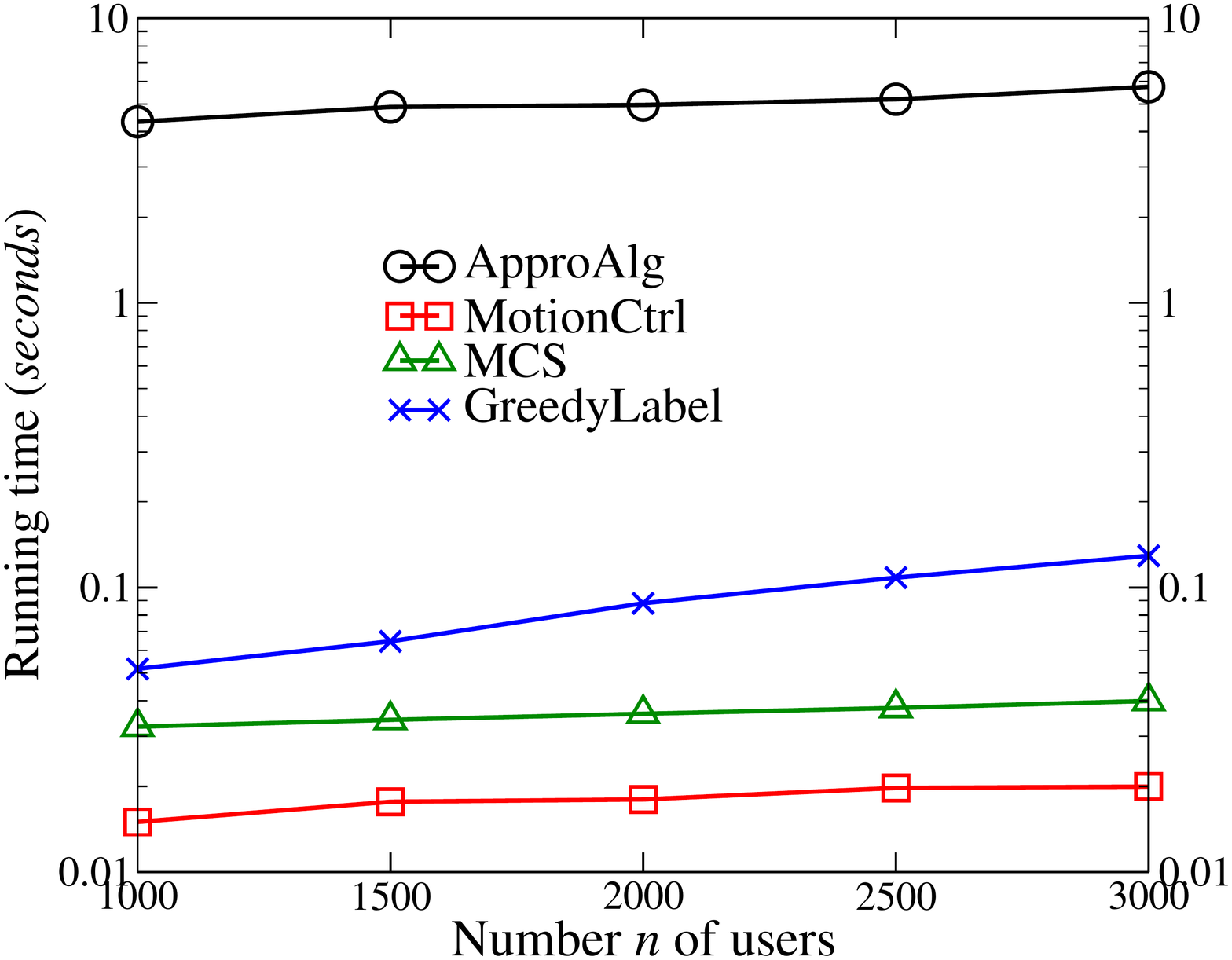}}
\vspace{-3mm}
\caption{The performance of different algorithms by increasing the number  of
  to-be-served users $n$ from 500 to 3,000, when the number of UAVs $K=30$.}
\label{vsNetwork}
\vspace{-3mm}
\end{figure*}

Following Lemma~\ref{lemmaSubmodular},
the constrained  maximum coverage problem can be cast as a submodular
function maximization problem, subject to a knapsack constraint.
Then, the algorithm in~\cite{S04} delivers a $(1-1/e)$-approximate solution $V_j$
to the constrained  maximum coverage problem with respect to node $v_j$, i.e.,
\begin{equation} \label{ratio1}
f(V_j \cup \{v_j\}) \ge (1-1/e)\cdot f(V^*_j \cup \{v_j\}),~\forall v_j \in V,
\end{equation}
assuming that $V^*_j$ is the optimal solution.

Assume that $v_l= \arg \max_{v_j \in V} \{ f(V^*_j \cup \{v_j\})\}$, i.e.,
 $f(V^*_l \cup \{v_l\})=\max_{v_j \in V} \{f(V^*_j \cup \{v_j\})\}$, where $1 \le l \le m$ and $m=|V|$.

Following {\tt Algorithm}~\ref{ApproAlg}, a set $S'_l$ of $K$ nodes is found
such that the induced graph $G[S'_l]$ is connected and the nodes in $V'_l
(=V_l\cup \{v_l\})$ are
contained in $G[S'_l]$.
We then have
{\small
\begin{eqnarray}
f(V') &=& \max_{v_j \in V} \{f(S'_j)\} ~\ge~f(S'_l)~\ge~ f(V_l\cup \{v_l\})\nonumber\\
   &&~~~\text{as $V_l\cup \{v_l\} \subseteq S'_l$ and $f(.)$ is nondecreasing} \nonumber\\
   &\ge& (1-1/e)\cdot f(V^*_l\cup \{v_l\}),~\text{by Ineq.~(\ref{ratio1})} \nonumber\\
   &=& (1-1/e)\cdot \max_{v_j \in V}\{ f(V_j \cup \{v_j\})\}\nonumber\\
   &\ge& \frac{1-1/e}{\lfloor \sqrt{K} \rfloor} \cdot OPT, ~\text{by Lemma~\ref{optUpperBound}.}
\end{eqnarray}
}

The time complexity of {\tt Algorithm}~\ref{ApproAlg} is analyzed as follows.
The running time of {\tt Algorithm}~\ref{ApproAlg} is dominated by
 Step~\ref{SetNotConnected}, which
finds a set $V_j$ of locations   for the  constrained maximum throughput problem with respect to each node $v_j$, by invoking the algorithm in~\cite{S04}.
Notice that the algorithm in~\cite{S04} invokes $O(Km^2)$ times of  {\tt Algorithm}~\ref{AlgMaxAssignment}.
Since the time complexity of is  {\tt Algorithm}~\ref{AlgMaxAssignment}
is $O((KC+n)^2 \log (KC+n))$ by Lemma~\ref{calfuncF},
the time complexity of {\tt Algorithm}~\ref{ApproAlg}
is $m\cdot O(K m^2) \cdot O((KC+n)^2 \log (KC+n))=O(K m^3 n^2 \log n)$,
since the maximum number $KC$ of users that can be served  by $K$ UAVs usually is in the order of the number $n$ of to-be-served users, i.e., $KC=O(n)$.
The theorem then follows.
\end{IEEEproof}

\section{Performance Evaluation} \label{secPerformance}
In this section, we evaluate the performance of the proposed algorithm.

\subsection{Experimental environment settings}
We consider a disaster area in a $3\times 3~km^2$ Euclidean space~\cite{ZWWW18}.
Assume that there are from 500 to 3,000 users  located in the disaster area,
and the human density follows the fat-tailed distribution,
that is,   most of people are crowded at a few places, while a small portion of people
are sparsely distributed at rest  places~\cite{SKWB10}.
Also, assume that the number  $K$ of to-be-deployed UAVs  is from 10 to 50.
Then, the approximation ratio of the proposed algorithm
is at least $\frac{1-1/e}{\lfloor \sqrt{50} \rfloor}=\frac{1-1/e}{7}$
following Theorem~\ref{theoremApproRatio},
where $e$ is the base of the natural logarithm.
Furthermore, the service capacity $C$ of each UAV varies from 50 users to 300 users.
In addition, assume that the $K$ UAVs hover at an altitude $h=300$~m~\cite{AKL14}.
The communication range  between  two UAVs is $R_{uav}=600$~m, whereas
 the communication range between a UAV and a ground user  is $R_{user}=500$~m~\cite{ZWWW18}.

The transmission power $P_t$ of each UAV is -6~dB,
the antenna gain $g_t$ is 5~dB,
and the noise power $P_N$ is -105~dB~\cite{YCX+19}.
Also, the transmission bandwidth $B_w$ is 180 kHz,
radio frequency $f_c$ is 2.5~GHz, and  speed $c$ of light
is $3\times 10^8$~m/s.
We consider an urban environment.
Then, the average shadow fadings for LoS and NLoS links are $\eta_{LoS}=1$~dB and $\eta_{NLoS}=20$~dB, respectively~\cite{AKL14}.
Finally, the LoS probability $p^{LoS}_{i,j}$ of a user $u_i$ served by a UAV
at location $v_j$ is $p^{LoS}_{i,j}=\frac{1}{1+a \cdot \exp(-b(\theta-a))}$,
where $a=9.611725$, $b=0.158062$, and $\theta$ is the elevation angle of user $u_i$ for the UAV deployed at $v_j$ at altitude $h$~\cite{AKL14}.

To evaluate the algorithm performance, we
 compare with  three state-of-the-art  algorithms as follows.
(i)  Algorithm {\tt MotionCtrl}~\cite{ZWWW18} delivers a  motion control solution to deploy $K$ UAVs to serve as many user as possible, while guaranteeing the UAV network connectivity.
(ii) Algorithm {\tt MCS}~\cite{KLT15} finds a $\frac{1-1/e}{5(\sqrt{K}+1)}$-approximate
solution to a problem of placing $K$ wireless routers so that
the value of a generalized submodular function over the placed routers is maximized, subject to the connectivity constraint.
(iii) Algorithm {\tt GreedyLabel}~\cite{KPK14} first
assigns a profit for
each node in a greedy strategy, then finds a connected  subgraph
with $K$ nodes such that the profit sum of the nodes  is maximized.
 All algorithms are implemented by the programming language C.
All  experiments are performed on a powerful server,
 which contains an Intel(R) Core(TM) i9-9900K CPU
 with 8 cores and each core having a maximum
turbo frequency of 5~GHz,
and 32 GB RAM.
 Notice that parallel computing is used.
The value  in each figure is the average of the results out
of 200 problem instances with the same network size.

\subsection{Algorithm Performance}

We first evaluated the algorithm performance  by increasing the number $n$ of users from 500 to 3,000, when there are $K=30$ UAVs in the network. Fig.~\ref{vsNetwork}(a) shows that the network throughput  by algorithm {\tt ApproAlg} is about from 10\% to 12\%
larger than those by algorithms {\tt MotionCtrl}, {\tt MCS}, and {\tt GreedyLabel}. For example, the network throughput by the four
algorithms {\tt ApproAlg}, {\tt MotionCtrl}, {\tt MCS}, and {\tt GreedyLabel}
are 25.6, 17.1, 23.1, and 18.4~Gbps, respectively  when there are 3,000
 users in the disaster area. Fig.~\ref{vsNetwork}(a)
 also demonstrates that the network
 throughput by each of the four algorithms becomes larger
 where there are more users in the network.
 Fig.~\ref{vsNetwork}(b) plots the average amount of energy
 consumed by the $K$ UAVs for flying
 from the service center in the disaster area to their service hovering locations,
 from which it can be seen that the average UAV flying energy consumption by each algorithm slightly becomes larger, as the $K$ UAVs need to be more sparsely deployed in a larger network, which incurs a longer flying distance.
Fig.~\ref{vsNetwork}(c) illustrates the running times of the four comparison algorithms. It can be seen that the running time of algorithm {\tt ApproAlg}
is around five seconds,  which is longer than those of the other three mentioned algorithms. Notice that such a short delay of a few seconds by algorithm {\tt ApproAlg}  is acceptable in real UAV networks, as the network throughput by the  algorithm is
up to 12\% larger than those by the other three algorithms. 

\begin{figure}[tp]
\vspace{-4mm}
\begin{center}
\includegraphics[scale=0.24]{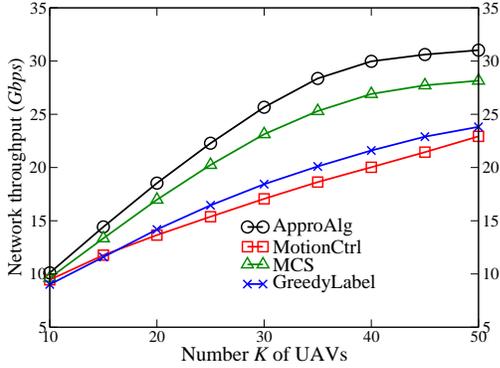}
\vspace{-5mm}
\caption{\small The performance of different algorithms by varying the number  of UAVs $K$ from 10 to 50, when  $n=3,000$  users.}
\label{vsNumUAVs}
\vspace{-7mm}
\end{center}
\end{figure}

We then studied the algorithm performance by varying the number $K$ of UAVs from 10 to 50 when there are $n=3,000$  users.
Fig.~\ref{vsNumUAVs} plots that the network throughput by each algorithm
increases with more UAV deployments, as more users can be served. In addition, the network throughput by algorithm {\tt ApproAlg} is from 5\% to 11\% higher than those by the other three algorithms. For example, the network throughput by the four algorithms {\tt ApproAlg}, {\tt MotionCtrl}, {\tt MCS}, and {\tt GreedyLabel}
 are 30.8,	20,	28.1, and 21.9~Gbps, respectively, when there are $K=40$ UAVs.

\begin{figure}[tp]
\begin{center}
\includegraphics[scale=0.24]{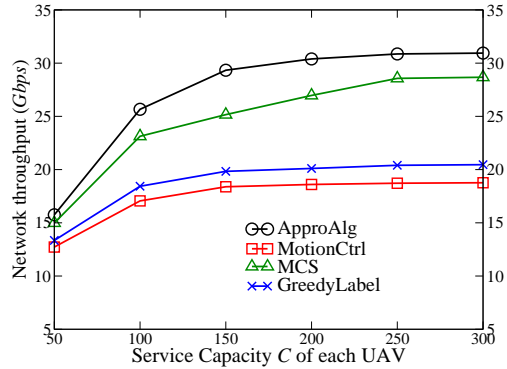}
\vspace{-5mm}
\caption{\small The performance of different algorithms by increasing the service capacity $C$ of each UAV from 50 users to 300 users,  when there are $n=3,000$
 users and $K=30$ UAVs.
 }
\label{vsCapUAV}
\end{center}
\vspace{-2mm}
\end{figure}

We thirdly investigated the performance of various algorithms by varying the service capacity $C$ of each UAV from 50 users to 300 users when there are $n=3,000$ users and $K=30$ UAVs in the monitoring area. Fig.~\ref{vsCapUAV} demonstrates
that the network throughput by each of the four algorithms increases with the growth of the service capacity $C$, as less numbers of UAVs are needed to serve the users in  places with high human densities and more UAVs thus can be used to serve the users in other places. Fig.~\ref{vsCapUAV}  shows that
the network throughput by each algorithm only slightly increases when
the service capacity $C$ of each UAV is larger than 150 users, as
there are a small portion of users located at other locations in the monitoring area, which needs several relaying UAVs to serve them~\cite{KPK14, LL05}.

\begin{figure}[tp]
\vspace{-3mm}
\begin{center}
\includegraphics[scale=0.24]{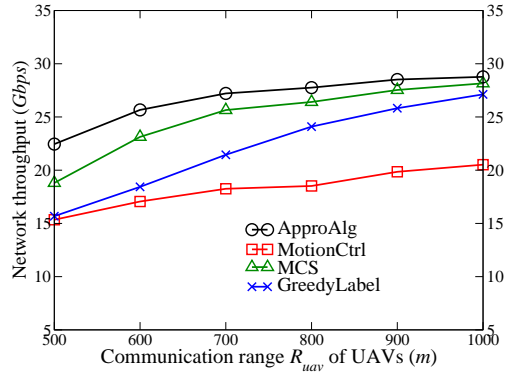}
\vspace{-5mm}
\caption{\small The performance of different algorithms by varying the
UAV communication range $R_{uav}$ from 500~m to 1,000~m while fixing $R_{user}=$500~m, when $n=3,000$,
$K=30$ and $C=100$.}
\label{vsCommUAVs}
\vspace{-5mm}
\end{center}
\end{figure}

We finally evaluated the algorithm performance by increasing the
UAV communication range $R_{uav}$ from 500~m to 1,000~m while fixing the user communication range $R_{user}$  at 500~m, when $n=3,000$,
$K=30$ and $C=100$. Fig.~\ref{vsCommUAVs} demonstrates that the network throughput by each of the four algorithms {\tt ApproAlg}, {\tt MotionCtrl}, {\tt MCS}, and {\tt GreedyLabel} increases with the growth of the UAV communication range $R_{uav}$. The rationale behind the phenomenon is that less numbers of relaying UAVs are needed when the UAV communication range $R_{uav}$ is larger, and more UAVs
thus can be used to serve users, thereby bringing about higher throughput.
Fig.~\ref{vsCommUAVs} also shows  the difference between the network throughput by algorithms {\tt ApproAlg},
{\tt MotionCtrl}, {\tt MCS}, and {\tt GreedyLabel}. For example, the network throughput by algorithm {\tt ApproAlg} is about 15\% larger than that by algorithm {\tt MCS}
when the UAV communication range $R_{uav}$ is 500~m, while
the network throughput by algorithm
{\tt ApproAlg} is only about 2.7\% higher than that by algorithm {\tt MCS}
when $R_{uav}=$1,000~m.

\section{Conclusions} \label{secCon}
In this paper, we studied the problem of deploying a communication network that consists of $K$ UAVs to provide temporarily emergent communications to people trapped in a disaster area. 
Under the assumption that the service capacity of each UAV is limited, and each of them only serves limited numbers of users, we investigated the problem of deploying $K$ UAVs as aerial base stations in the top of a disaster area, such that the sum of the data rates of users served by the UAVs is maximized, subject to that (i) the number of users served by each UAV is no greater than its service capacity, and (ii) the communication network induced by the $K$ UAVs is connected. We devised a novel $\frac{1-1/e}{\lfloor \sqrt{K} \rfloor}$-approximation algorithm for the problem, where $e$ is the base of the natural logarithm. We also evaluated the  algorithm performance via simulation experiments. Experimental results showed that the proposed algorithm is very promising. Especially, the network throughput by the proposed algorithm is up to 12\% higher than those by existing algorithms.

\section*{Acknowledgement}	
The work by Wenzheng Xu was supported by the National Natural Science Foundation of China (NSFC) with grant number 61602330, Sichuan Science and Technology Program (Grant No. 2018GZDZX0010 and 2017GZDZX0003), and the National Key Research and Development Program of China (Grant No. 2017YFB0202403). The work by Weifa Liang was supported by Australian Research Council under its Discovery Project Scheme with Grant No. DP200101985.

\vspace{-1.5cm}
\begin{IEEEbiography}
[{\includegraphics[width=1.0in,height=1.25in,clip, keepaspectratio]{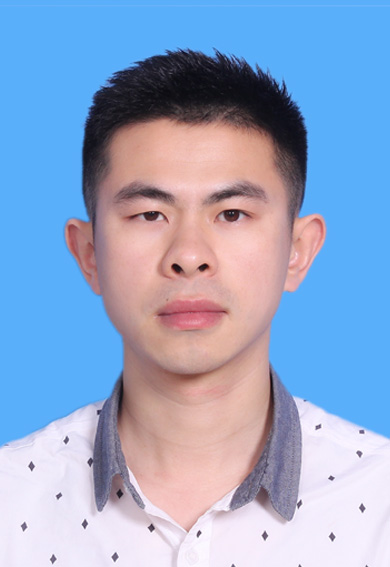}}]
{Wenzheng Xu} (M'15) received the BSc, ME, and PhD degrees in computer science from Sun
Yat-Sen University, Guangzhou, P.R. China, in 2008, 2010, and 2015, respectively.
He currently is an Associate Professor at Sichuan University. Also, he was a visitor at both the Australian National University and the Chinese University of Hong Kong.
His research interests include wireless ad hoc and sensor networks, mobile computing, approximation algorithms, combinatorial optimization, online social networks, and graph theory. He is a member of the IEEE.
\end{IEEEbiography}

\vspace{-1.5cm}
\begin{IEEEbiography}
[{\includegraphics[width=1.0in,height=1.25in,clip, keepaspectratio]{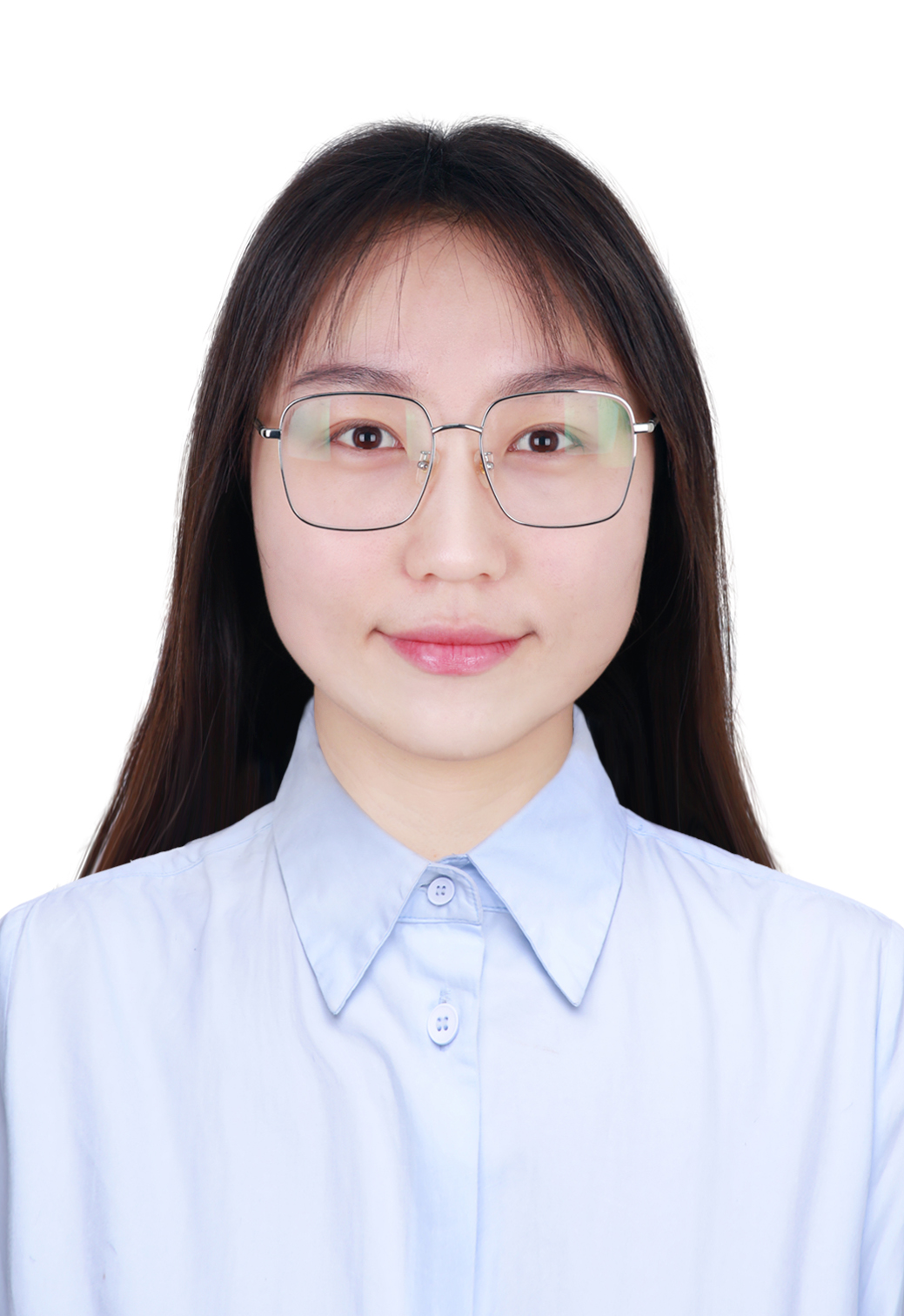}}]
{Yueying Sun}  received the BSc degree in computer science from Henan University,
P. R. China, in 2019. She now is a second year
master student in computer science at Sichuan
University. Her research interests include UAV
 networking and mobile computing.
\end{IEEEbiography}

\vspace{-1.5cm}
\begin{IEEEbiography}
[{\includegraphics[width=1.0in,height=1.25in,clip, keepaspectratio]{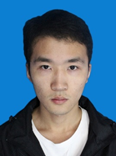}}]
{Rui Zou}  received the BSc degree in electronical  science and technology from Wuhan University of Technology,
P. R. China, in 2018. He now is a third year
master student in computer science at Sichuan
University. His research interests include wireless
 networks and mobile computing.
\end{IEEEbiography}

\vspace{-1.6cm}
\begin{IEEEbiography}
[{\includegraphics[width=1in,height=1.25in,clip,keepaspectratio]{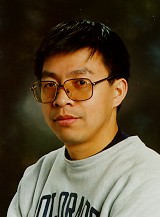}}]
{Weifa Liang} (M'99--SM'01) received the PhD degree from the Australian
National University in 1998, the ME degree from the University of
Science and Technology of China in 1989, and the BSc degree from Wuhan
University, China in 1984, all in Computer Science. He currently  is a
Professor in the Department of Computer Science at City University of
Hong Kong. Prior to the current position, he was a Professor at the Australian National University. His
research interests include design and analysis of energy efficient
routing protocols for wireless ad hoc and sensor networks, Internet of
Things, edge and cloud computing, Network Function Virtualization and
Software-Defined Networking, design and analysis of parallel and
distributed algorithms, approximation algorithms, combinatorial
optimization, and graph theory. He serves as an Associate Editor for the
IEEE Trans. Communications. He is a senior member of the IEEE.
\end{IEEEbiography}

\vspace{-13mm}
\begin{IEEEbiography}[{\includegraphics[width=1in,height=1.25in,clip,keepaspectratio]
{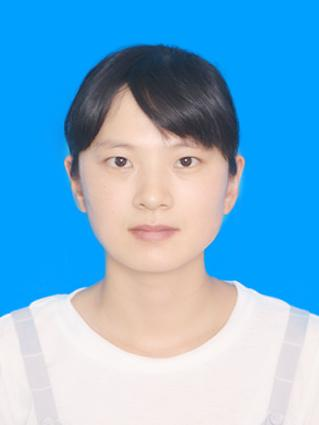}}]{Qiufen Xia} received her PhD degree from the Australian National University in 2017, the ME degree and BSc degree from Dalian University of Technology (DUT) in China in 2012 and 2009, all in Computer Science. She is currently a lecturer in the International School of Information Science and Engineering at DUT. Her research interests include mobile cloud computing, query evaluation, big data analytics, big data management in distributed clouds, and cloud computing.
\end{IEEEbiography}

\vspace{-15mm}
\begin{IEEEbiography}[{\includegraphics[width=1in,height=1.25in,clip,keepaspectratio]
{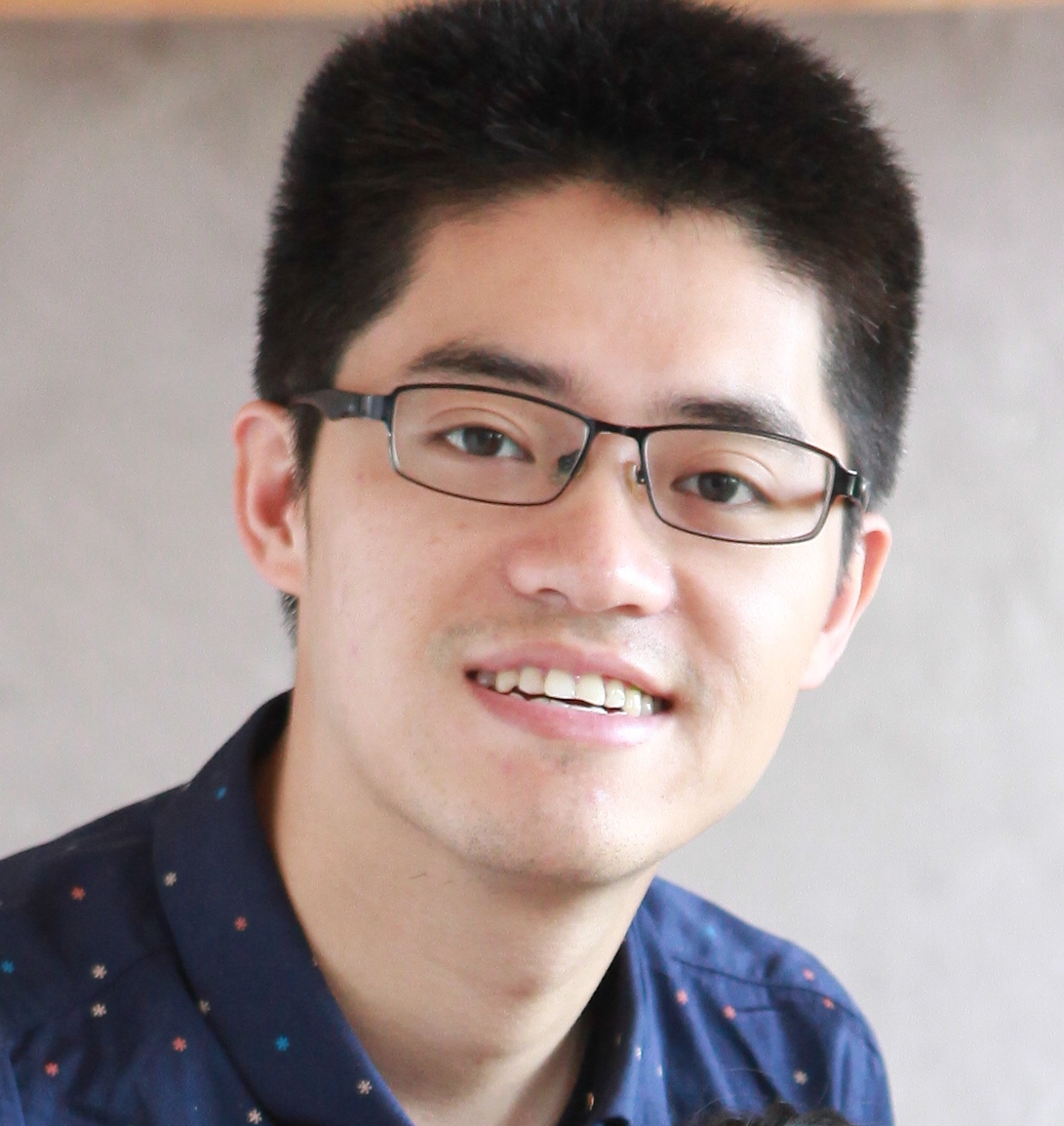}}]{ Feng Shan} (M'17) received the Ph.D. degree
in computer science from Southeast University,
Nanjing, China, in 2015.
He is currently an Assistant Professor with
the School of Computer Science and Engineering,
Southeast University. He was a Visiting Scholar
with the School of Computing and Engineering,
University of Missouri-Kansas City, Kansas City,
MO, USA, from 2010 to 2012. His current research
interests include energy harvesting, wireless power
transfer, swarm intelligence, and algorithm design
and analysis.
\end{IEEEbiography}

\vspace{-1.6cm}
\begin{IEEEbiography}
[{\includegraphics[width=1in,height=1.25in,clip,keepaspectratio]{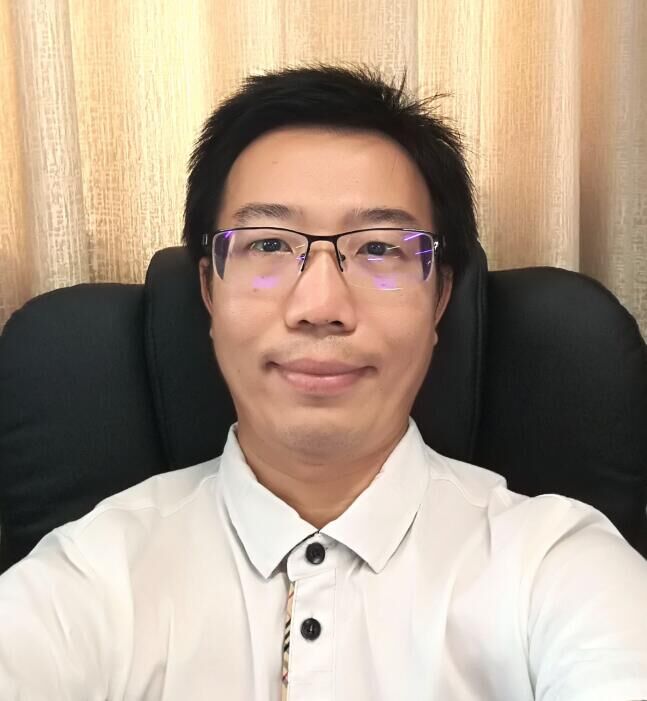}}]
{Tian Wang} received his BSc and MSc degrees in Computer Science from the Central South University in 2004 and 2007, respectively. He received his PhD degree in City University of Hong Kong in 2011. Currently, he is a professor in the National Huaqiao University of China. His research interests include wireless sensor networks, social networks and mobile computing.
\end{IEEEbiography}

\vspace{-15mm} \begin{IEEEbiography}[{\includegraphics[width=1in,height=1.25in,clip,keepaspectratio]{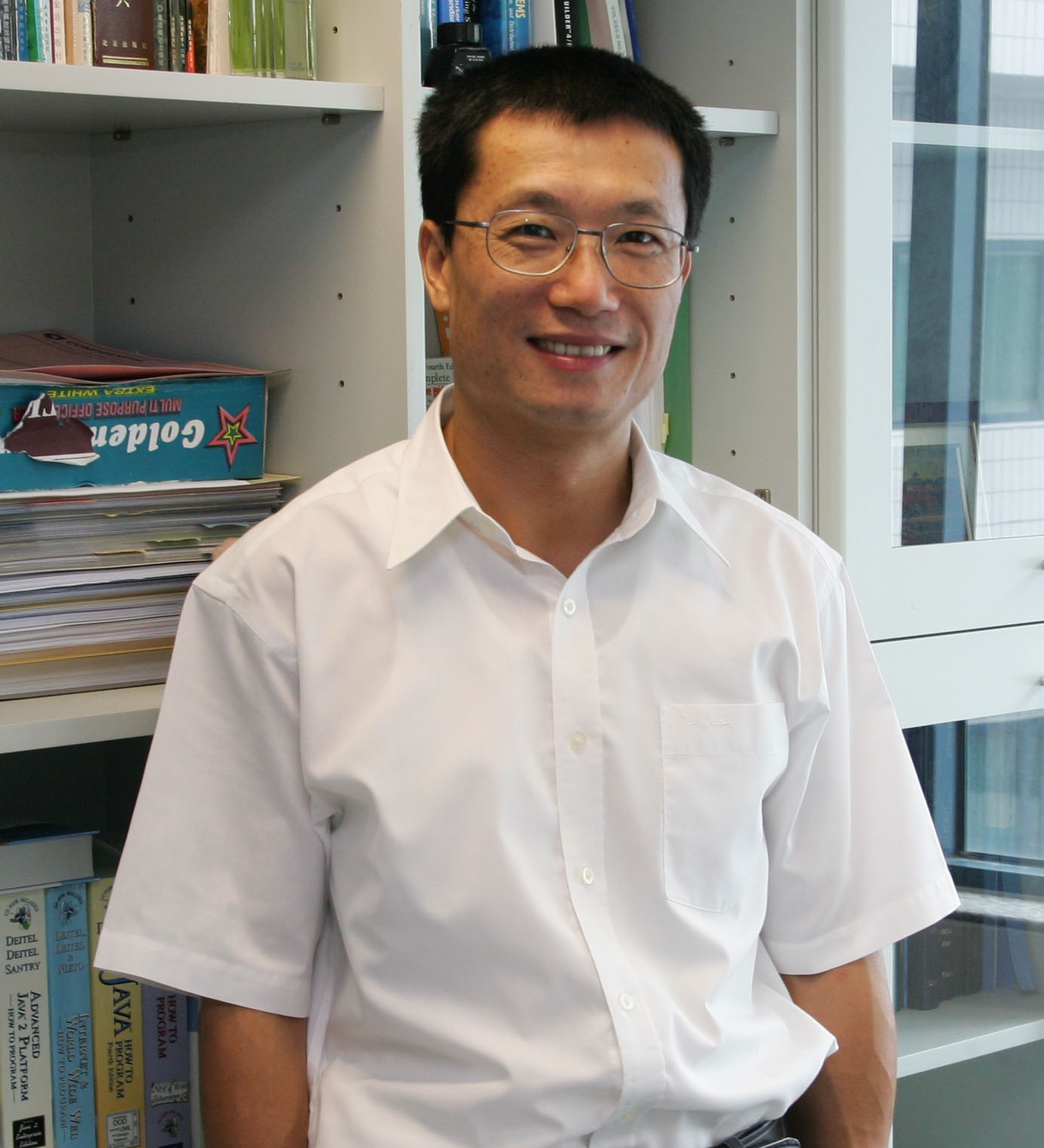}}] {Xiaohua Jia} (A'00--SM'01--F'13) received the BSc and MEng degrees in 1984 and 1987, respectively, from the University of Science and Technology of China, and DSc in 1991 in information science from the University of Tokyo. He is currently a chair professor with Department of Computer Science at City University of Hong Kong. His research interests include cloud computing and distributed systems, computer networks, wireless sensor networks and mobile wireless networks.
He is an editor of IEEE Transactions on Parallel and Distributed Systems (2006-2009),  Journal of World Wide Web, Wireless Networks, Journal of Combinatorial Optimization,
and so on. He is the general chair of ACM MobiHoc 2008, TPC co-chair of IEEE MASS 2009, area-chair of IEEE INFOCOM 2010, TPC co-chair of IEEE GlobeCom 2010, Ad Hoc and Sensor Networking Symposium,
and Panel co-chair of IEEE INFOCOM 2011. He is a fellow of the IEEE.
\end{IEEEbiography}

\vspace{-1.2cm}
\begin{IEEEbiography}
[{\includegraphics[width=1in,height=1.25in,clip,keepaspectratio]{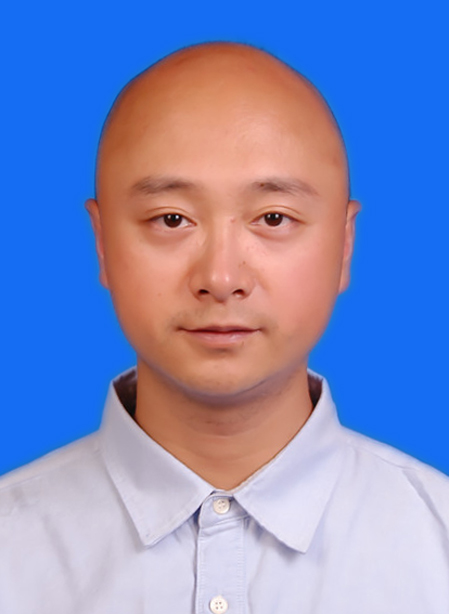}}]
{Zheng Li} received the BSc degree in computer science in 1997, the ME degree in computer science in 2000, and the PhD in Applied Mathematics in 2009, all from Sichuan University.
He currently is a full professor at Sichuan University.
His research interests include computer vision, approximation algorithms, wireless sensor networks,   combinatorial optimization,  and parallel and distributed algorithms.
\end{IEEEbiography}

\end{document}